\documentclass[journal]{new-aiaa}

\usepackage[utf8]{inputenc}
\usepackage{textcomp}

\usepackage{graphicx}
\usepackage{amsmath}
\usepackage[version=4]{mhchem}
\usepackage{siunitx}
\usepackage{longtable,tabularx}

\usepackage{color}
\usepackage{amsmath} 
\usepackage{cite}
\usepackage{mathtools}
\usepackage{array}
\usepackage{algpseudocode,algorithm,algorithmicx}
\usepackage{color}
\usepackage{booktabs}
\usepackage[hyphens]{url}
\usepackage{lipsum}
 \usepackage[dvipsnames]{xcolor}
\usepackage[deletedmarkup=sout,authormarkup=superscript]{changes}
\usepackage[normalem]{ulem}
\usepackage{flushend}
\usepackage{svg}

\newcommand{\R}{\mathbb{R}}

\newcommand{\reals}{\mathbb{R}}

\definechangesauthor[name={Justin}, color=NavyBlue]{JK}
\definechangesauthor[name={Dominic}, color=RedViolet]{DLM}
\definechangesauthor[name={Chikeng}, color=OliveGreen]{CKD}

\graphicspath{ {./figures/} }
\usepackage{xcolor}
\usepackage{soul}
\newcommand{\change}[1]{#1}

\title{Adaptive Long-range UAV Flight Planning using Monte Carlo Search Trees}

\author{Chi Keng Dong \footnote{Graduate Research Assistant, Dept. of Mechanical Engineering, University of British Columbia, Vancouver, BC, Canada. Email: dzjapple@outlook.com}, Justin Kernot\footnote{Graduate Research Assistant, Dept. of Mechanical Engineering, University of British Columbia, Vancouver, BC, Canada. Email: jkernot@student.ubc.ca}, and Dominic Liao-McPherson\footnote{Assistant Professor, Dept. of Mechanical Engineering, 
University of British Columbia, Vancouver, BC, Canada. Email: dliaomcp@mech.ubc.ca}}
\affil{University of British Columbia, 2054-6250 Applied Science Lane, Vancouver BC, V6T 1Z4, Canada}

\begin{document}

\maketitle

\begingroup\small\noindent
\textit{Authors' accepted manuscript. Copyright \copyright{} American Institute of Aeronautics and Astronautics. The version of record is published by AIAA, DOI: \url{https://doi.org/10.2514/1.G010042}.}
\endgroup
\medskip

\begin{abstract}
Long-range fixed-wing UAV missions require energy-efficient, dynamically feasible trajectories that can adapt online to evolving wind and airspace constraints. We present a Monte Carlo Tree Search (MCTS)–based framework that addresses this challenge by formulating mission planning as a finite-horizon Markov decision process incorporating aircraft kinematics, onboard energy dynamics, forecast wind fields, terrain, and time-varying no-fly zones. The MCTS planner searches over waypoint references generated by pre-stabilizing feedback controllers and a constraint-aware navigation field, yielding dynamically feasible trajectories. Simulation results from medical delivery scenarios show that the MCTS planner outperforms baseline \change{geometric planners and population-based optimization algorithms.} Moreover, the MCTS planner can be used in closed-loop to rapidly computes safe diversion trajectories under evolving constraints.
\end{abstract}

\section*{Nomenclature}

{\renewcommand\arraystretch{1.0}
\noindent\begin{longtable*}{@{}l @{\quad=\quad} l@{}}
$X(t)$ & aircraft state vector \\
$\psi(t)$ & heading angle \\
$E_n(t)$ & normalized onboard energy \\
$p(t)$ & spatial position vector \\
$U(t)$ & control input vector \\
$v_a(t)$ & air speed of the aircraft \\
$\gamma(t)$ & flight path angle \\
$\omega(t)$ & heading rate \\
$w(x,y,z,t)$ & wind field vector \\
$P(X,U)$ & instantaneous power draw \\
$D$ & total drag \\
$m$ & aircraft mass \\
$g$ & gravitational acceleration \\
$\eta$ & propulsion efficiency \\
$P_0$ & baseline power draw \\
$\rho(z)$ & air density \\
$A$ & wing surface area \\
$C_D$ & drag coefficient \\
$C_{D0}$ & zero-lift drag coefficient \\
$K$ & induced drag factor \\
$C_L$ & lift coefficient \\
$\rho_0$ & sea-level air density \\
$T_0$ & sea-level temperature \\
$k$ & temperature lapse rate \\
$R$ & gas constant \\
$\varepsilon_t$ & target region radius \\
$P_T$ & target region \\
$T$ & mission completion time \\
$\Omega$ & unsafe states \\
$\Omega_n(t)$ & no-fly zones \\
$\varepsilon_z$ & minimum altitude margin \\
$z_h$ & elevation \\
$x_r, y_r$ & surface position \\
$z_{\max}$ & altitude ceiling \\
\end{longtable*}}
\newpage

\section{Introduction}
(Semi-)autonomous long-range fixed-wing uncrewed aerial vehicles (UAVs) are increasingly being deployed across a wide range of applications. UAVs have enabled life-saving medical supply delivery to remote and resource-constrained regions in African countries \citep{griffith2023rvf,amukele2022using,snouffer2022six}, maritime surveillance and inland water monitoring \citep{eu2025maritime,temlin2023inland}, wildfire monitoring and prevention \citep{patrinopoulou2024distributed}, and geological mapping and remote sensing \citep{everaerts2008uav,nex2014uav}.

Across these diverse missions, a core challenge is \textit{motion planning}, generating feasible, safe, and efficient trajectories over long distances while respecting nonholonomic dynamics, restricted airspace, tight energy budgets, and environmental disturbances. In autonomous flight, \textit{re-planning} is essential to ensure robustness and adaptability during execution, since unreliable communication and evolving environmental conditions (e.g., emerging no-fly zones, unexpected obstacles, and changing weather patterns) can render pre-flight plans invalid. In such cases, UAVs must be able to re-plan the flight path in real time using onboard computing resources. This challenge is compounded by stringent regulatory frameworks in many jurisdictions, where practical methods must be interpretable, provably safe, and highly reliable \citep{transport2024rules}. Moreover, many UAV platforms must meet strict size, weight, and power constraints, which restrict the onboard computational resources available. Together, these factors motivate the need for online re-planning frameworks that are computationally efficient, interpretable, and trustworthy in safety-critical long-range operations. 

Several methods have been applied to solve the motion planning problem for fixed-wing UAVs including Graph- and grid-based methods such as A*, Theta*, wavefront expansion, and Dijkstra's Algorithm \citep{hwangbo2007efficient,bassolillo2024path,medeiros2010dijkstra,mothes2019trajectory}. \change{These methods can readily incorporate energy into planning \citep{ChakrabartyA_2009,ChakrabartyA_2010,ChakrabartyA_2011} and have been shown to be computationally efficient while providing completeness guarantees. However, they require an a-priori discretization of the domain and suffer from discretization errors and the curse of dimensionality.} Sampling-based methods such as Rapidly-Exploring Random Trees (RRTs) \citep{lavalle2001randomized} and their real-time variants like RT-RRT* \citep{naderi2015rt} and RRTx \citep{otte2016rrtx} bypass these limitations through random exploration of the state space and have been applied extensively to fixed-wing UAVs; examples include integration with guidance laws to generate dynamically feasible paths \citep{ramana2016motion}, kino-dynamic RRT variants that embed dynamics directly into the search \citep{yavari2020lazy,ge2022trajectory,ChakrabartyA_2013}, and motion-primitive–based approaches using trajectory libraries \citep{lee2018rrt,allen2016toward,levin2018motion,hwangbo2007efficient,dallalbashi2024fixedwing3drrt}. These methods represent the state of the art in nonconvex environments and are excellent path planners, but they remain limited by their formulations as two-point boundary value problems, which can make it challenging to handle motion planning problems with objectives beyond minimum distance path planning.

Trajectory optimization (TO) methods have also been widely applied to UAV motion planning tasks. For agile collision-avoidance manoeuvres, mixed integer linear programming and nonlinear model predictive control have been used to generate dynamically feasible trajectories in cluttered environments \citep{zhang2018collision,basescu2020direct,mothes2019trajectory,samaniego2019smooth}. For long-range navigation under obstacle constraints, sequential quadratic programming and sequential convex programming have been explored to balance efficiency and feasibility in extended missions \citep{bertoncini2014fixed,schopferer2018path,bassolillo2024path}. Early studies also demonstrated the feasibility of TO frameworks for aircraft and UAV applications more broadly \citep{richards2002aircraft,medeiros2010dijkstra}. TO methods are generally scalable with horizon length, but are sensitive to local minima and can be unreliable in nonconvex environments. Global optimization methods such as Genetic Algorithms (GA)~\citep{goldberg1989genetic} and Particle Swarm Optimization (PSO)~\citep{kennedy1995particle} can avoid local minima and are effective for offline pre-flight planning \citep{hovenburg2020longrange,li2006pso,roberge2018fast}, but their computational burden makes them unsuitable for online re-planning in long-range UAV missions.

Artificial Intelligence (AI)–based approaches have gained traction in recent years. Reinforcement Learning (RL)–based policies, typically parameterized by neural networks (NNs), have been applied to UAV path planning and obstacle avoidance in dynamic environments \citep{zhuang2021optimization,zhao2018q,xu2020neural}. These methods can be fast and performant, but they are generally unreliable, uninterpretable, and demand massive training datasets, making them largely unsuitable for safety-critical domains.

Finally, sampling-based Markov Decision Process (MDP) methods, notably Monte Carlo Tree Search (MCTS) \citep{chaslot2008monte,kolobov2012planning,browne2012survey} and Model Predictive Path Integral Control (MPPI) \citep{kazim2024path} formulate mission planning as a sequential decision-making problem under uncertainty, which is then solved using stochastic sampling-based algorithms. Their formulation as a stochastic optimal control problem naturally accommodates rich mission objectives and they can be readily integrated into re-planning schemes using the model predictive control framework~\citep{rawlings2009model}. \change{E.g., ad-hoc methods have been applied successfully to soaring flight \citep{LangelaanJW_2008}}, MPPI has been used for short-horizon, reactive navigation of agile UAVs in cluttered environments \citep{minavrik2024model,higgins2023model}, and MCTS has been applied to a wide range of decision-making and control problems \citep{swiechowski2021monte,kemmerling2024beyond}, demonstrating significant potential for re-planning in quadrotor and spacecraft applications \citep{riviere2021spectral,lathrop2023mpt,qian2022mcts}.

\change{In this work, we develop an energy- and wind-aware MCTS planner for long-range flight that supports fast online re-planning in dynamic environments.} Our algorithm incorporates long-range kinematics and energy consumption models, and plans in a reference space induced by a set of feedback controllers that emulate the UAV’s low-level behaviour, including basic terrain and collision avoidance. This results in a computationally efficient and kinematically compatible motion planner that is reliable in dynamic and nonconvex environments, avoiding the curse-of-dimensionality issues common in RRT-based kino-dynamic planners and the local minima issues of gradient-based trajectory optimization approaches. We demonstrate the capability of the proposed planner through simulations, showing that it generates energy-efficient, wind-aware trajectories over long distances, and performs fast online re-planning when confronted with emerging environmental constraints.

The rest of the paper is organized as follows: 
Section~\ref{sec:problem_formulation} defines the problem setup and introduces the aircraft dynamics and energy models. 
Section~\ref{sec:planning_algorithm} presents the overall planning framework, including the Open-loop planner (Section~\ref{ssec:openloop_planner}) with MDP reformulation and the MCTS algorithm, and the Closed-loop re-planner (Section~\ref{ssec:closedloop_re-planning}) by solving the MCTS algorithm repeatedly. Simulation results and performance evaluations are provided in Section~\ref{sec:simulation_results}, followed by concluding remarks in Section~\ref{sec:conclusion}. 

\section{Problem Formulation}
\label{sec:problem_formulation}
This section formulates the mission planning problem for a long-range fixed-wing UAV as a trajectory optimization problem with the goal of reaching a target while minimizing energy usage and avoiding terrain and no-fly zones.

\subsection{Aircraft Model}  \label{ssec:aircraftEnvModel}
 
\begin{figure}[h!]
    \centering
    \includegraphics[width=0.48\textwidth]{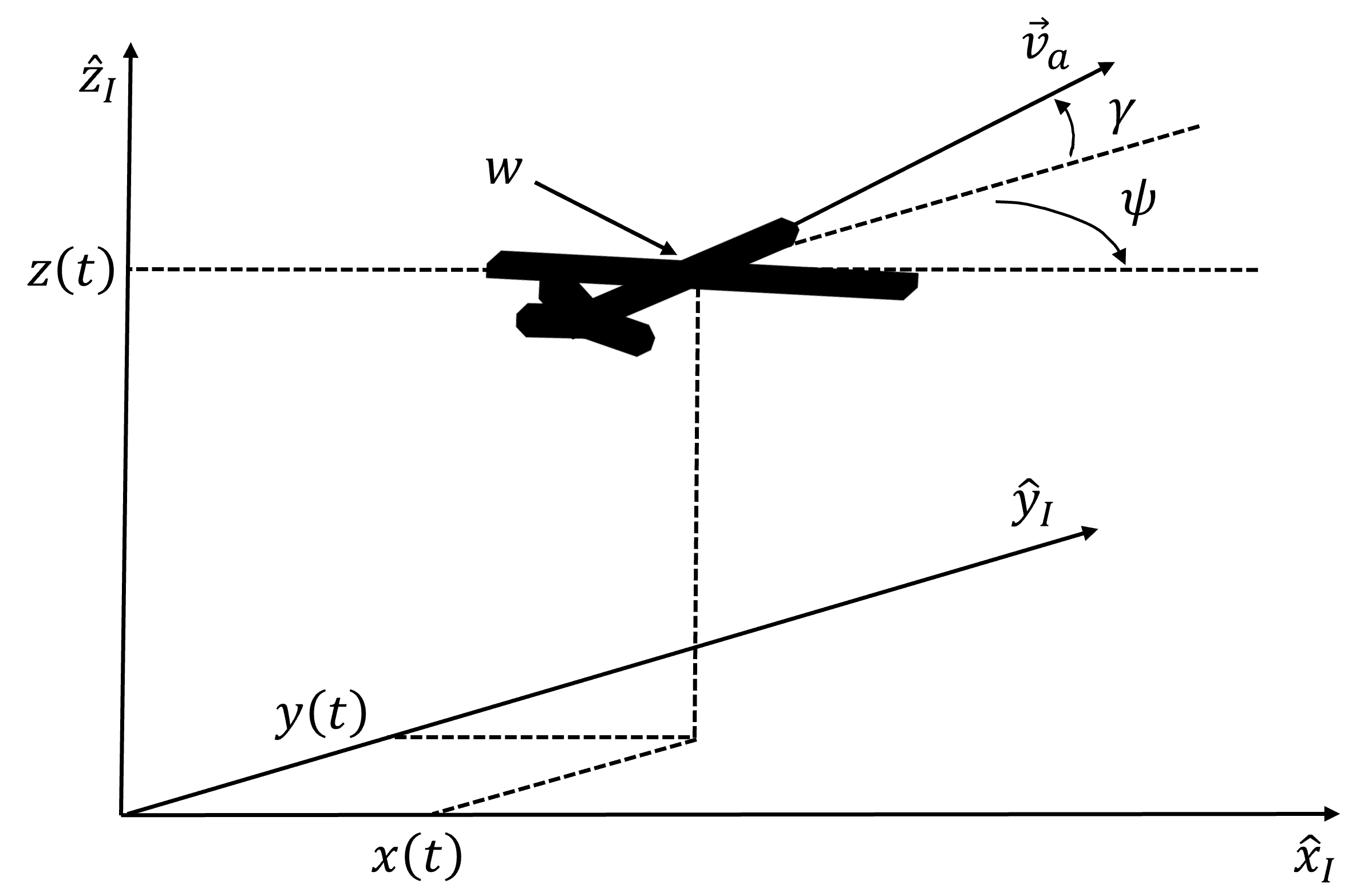}
    \caption{Kinematic states of the aircraft relative to the world frame, with $x(t)$ and $y(t)$ being the surface position, $z(t)$ the altitude above sea level, $\psi$ the heading angle, and $\gamma$ the flight path angle. The world frame $xy$-plane is tangent to the Earth's surface at sea level, with $\hat x_I$ pointing east and the $\hat y_I$ pointing north. \change{The wind speed vector $w=[w_x~~w_y~~w_z]^\top$ acts on the aircraft as an external disturbance.}}
    \label{fig:aircraftPose}
\end{figure}

For the aircraft model, we use a coupled four-degree-of-freedom (4-DOF) kinematic and energy consumption model. The state of the system is $X = [x~y~z~\psi~E_{n}]^\top$, where $p=[x~~y~~z]^\top$ is the position of the aircraft with respect to an inertial reference frame, $\psi$ is the heading angle, and $E_{n}$ is the normalized onboard energy. Figure~\ref{fig:aircraftPose} illustrates the kinematic states of the aircraft relative to the world frame.

The admissible state space is defined by Eq.~\ref{eq:admiss}, where the UAV must maintain a minimum altitude margin $\varepsilon_z$ above the terrain elevation $z_h = \operatorname{f}(x, y)$, while remaining below the prescribed altitude ceiling $z_{\max}$ above mean sea level. 

\begin{equation} \label{eq:admiss}
    \mathcal{X} = \left\{ [x~y~z~\psi~E_{n}]^\top~\middle|~ z \ge \operatorname{f}(x,y) + \varepsilon_z,\; z \le z_{\max},\; E_n \in [0,1],\; \psi \in (-\pi,\pi] \right\}
\end{equation}

The control input is $U = [v_a~\gamma~\omega]^\top$, where the aircraft is controlled by instantaneous and direct actuation to the 
\change{airspeed} $v_a(t)\in\R$, the flight path angle $\gamma(t)\in (-\pi,\pi]$, and the heading rate $\omega(t)\in\R$. In addition to spatial constraints, the control inputs are bounded by aircraft performance limits. The set of admissible controls is

\begin{equation}
\label{eq:inputConstraints}
\mathcal{U} =
\left\{
[v_a,\,\gamma,\,\omega]^\top \in \mathbb{R}^3 ~\bigg|~
v_{a,\min} \le v_a \le v_{a,\max},\;
\gamma_{\min} \le \gamma \le \gamma_{\max},\;
\omega_{\min} \le \omega \le \omega_{\max}
\right\}
\end{equation}

The aircraft is powered by an electric propulsion system with finite energy capacity and is subject to an external wind field $w(x, y, z, t) = [w_x~w_y~w_z]^\top \in \mathbb{R}^3$. Approximating the aircraft with a modified unicycle model \citep{lee2000stable}, the kinematic and normalized onboard energy dynamics are

\begin{subequations}
\begin{align}
    \dot{x}&=v_a\cos \gamma \cos \psi + w_x\left(x,y,z,t \right ) \label{eq:kin_x} \\
    \dot{y}&=v_a\cos \gamma \sin \psi + w_y\left(x,y,z,t \right ) \\
    \dot{z}&=v_a \sin \gamma + w_z\left(x,y,z,t \right )\\
    \dot{\psi}&=\omega \label{eq:kin_psi}\\
    \dot{E}_{n}&= -\frac{P(X,U)}{E_{\text{max}}}
    \label{eq:E}
\end{align}
\end{subequations}

\noindent where $P(X,U)$ is the instantaneous power draw and $E_{\text{max}}$ is the maximum stored energy. \change{Unicycle-type models are commonly used in long-range fixed-wing UAV motion planning \citep{heintz2020formation,ren2003clf,he2023roto,LangelaanJW_2008,ChakrabartyA_2011} where flight dynamics are less critical since they are low-complexity but still capture the essential nonholonomic kinematics of fixed-wing aircraft. For our purposes, we chose a 4-DOF unicycle model to account for the aircraft's spatial position and heading angle by Eqs.~\ref{eq:kin_x}-\ref{eq:kin_psi}.} The total power consumption, accounting for aerodynamic drag and climb/descent rate, is modelled as

\change{
\begin{equation}
\label{eq:power}
  P = P_0 + \max\{\eta^{-1}(v_a D + mg\dot{z}),0\}
\end{equation}}

\noindent where $D$ is the total drag, the term $v_a D$ is the power required to overcome aerodynamic drag, $m$ is the aircraft mass, $g$ is the gravitational acceleration, and $m g\dot{z}$ is the power required to overcome gravity. The parameter $\eta \in [0,1]$ denotes the propulsion efficiency, and $P_0$ is the baseline power draw due to avionics, leakage, and other fixed loads. \change{The $\max\{\cdot,0\}$ term encodes that the aircraft cannot regenerate power during a descent.} The aerodynamic drag $D$ obeys the standard drag polar \citep{anderson1999aircraft} with $D(v_a,z) = \frac12 \rho v_a^2 A C_D,\quad C_D = C_{D0} + K C_L^2,\quad C_L = \frac{2mg\cos\gamma}{\rho v_a^2 A}$, where $\rho$ is the air density, $A$ is the reference area, $C_D$ is the drag coefficient, $C_{D0}$ is the zero-lift drag coefficient, $K$ is the induced drag factor, and $C_L$ is the lift coefficient. The dependence of air density on altitude $\rho(z)$ is modelled using the International Standard Atmosphere (ISA) model, $\rho(z) = \rho_0 \left(1 + \frac{k z}{T_0} \right)^{-\left(1 + \frac{g}{Rk} \right)}$, where $\rho_0$ is the sea-level density, $T_0$ is the reference temperature, $k$ is the temperature lapse rate, and $R$ is the specific gas constant for air.

Combining Eqs.~\eqref{eq:kin_x}–\eqref{eq:E}, the continuous-time dynamics can be compactly written as

\begin{equation}
    \dot{X} = \mathcal{F}(X,U,w) = [\dot x~\dot y~\dot z~\dot \psi~\dot E_n]^\top
    \label{eq:dynamics}
\end{equation}

\subsection{Motion Planning Problem}
\label{ssec:motionPlanningProblem}

The objective of the motion planning problem is to find a feasible, energy-efficient trajectory from the initial spatial position of the UAV ${p}_s \in \reals^3$ to a target region $P_T \coloneqq \{p \in \reals^3 : \|p-p_t\| \leq \varepsilon_t\}$ centered at ${p}_t \in \reals^3$ with radius $\varepsilon_t \in \R$. The environment includes complex terrain, no-fly zones, and space- and time-varying wind fields. We pose the motion planning problem as a finite-horizon optimal control problem:

\begin{subequations}\label{eq:constrained}
\begin{align}
\max_{X(\cdot),~U(\cdot)} \quad & E\big(T\big) \\[3pt]
\text{s.t.} \quad & \dot{X}(t) = \mathcal{F}(X(t) , U(t),w(t)), \label{eq:transition_equation} \quad \forall t \in [0, T]\\[3pt]
& X(0) = \xi, \change{ \quad X(T) \in \big(P_T\times(-\pi,\pi]\times[0,1]\big)}\\[3pt]
& U(t) \in \mathcal{U}, \quad X(t) \in \mathcal{X}, \quad h(X(t)) \notin \Omega_n(t)
\end{align}
\end{subequations}

\noindent where $\xi = [p_s~\psi_0~E_{n,0}]^\top$ is the initial state of the aircraft, $\mathcal{U}$ is the set of admissible inputs, $w(t)$ represents the wind field forecast, and $\Omega_n(t)$ denotes the set of unsafe states of time-varying no-fly zones. We also define the map $p = h(X)$ which extracts the spatial position $p=[x~~y~~z]^\top$ from the state.

The no-fly zones represent e.g., $\Omega_n(t)$, are modelled as infinitely tall cylindrical regions with vertical axes aligned with the inertial $z$ axis. Each no-fly zone is defined by a center point $(x_{c_i}(t), y_{c_i}(t))$ and a radius $R_{c_i}(t)$ for $i = 1,\dots,N_c(t)$ at time $t$, where $N_c(t)$ is the total number of active no-fly zones. The spatial constraint for each no-fly zone at time $t$ is $(x(t) - x_{c_i}(t))^2 + (y(t) - y_{c_i}(t))^2 \leq R_{c_i}(t)^2$, and the no-fly zone region is defined by the union of all no-fly cylinders at time $t$,

\begin{equation} \label{eq:nfzUnion}
    \Omega_{\text{n}}(t) = \bigcup_{i=1}^{N_c(t)} \left\{ [x~~y~~z]^\top \in \R^3 \;\middle|\; (x - x_{c_i})^2 + (y - y_{c_i})^2 \leq R_{c_i}^2 \right\}
\end{equation}

\begin{figure}[h!]
    \centering
    \includegraphics[width=.5\textwidth]{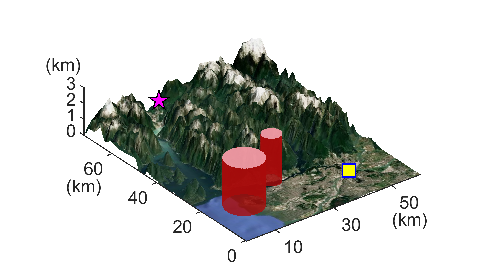}
    \caption{Visual representation of the terrain and no-fly zone topography.}
    \label{fig:terrain}
\end{figure}

A visual representation of the environment and constraint topology is shown in Fig.~\ref{fig:terrain}, where the axes represent ground-plane position and elevation in meters above sea level. The initial UAV position (yellow square) and target region center (magenta star) are marked.

\section{Planning Algorithm}
\label{sec:planning_algorithm}

\begin{figure}[h!]
    \centering
    \includegraphics[width=0.5\textwidth]{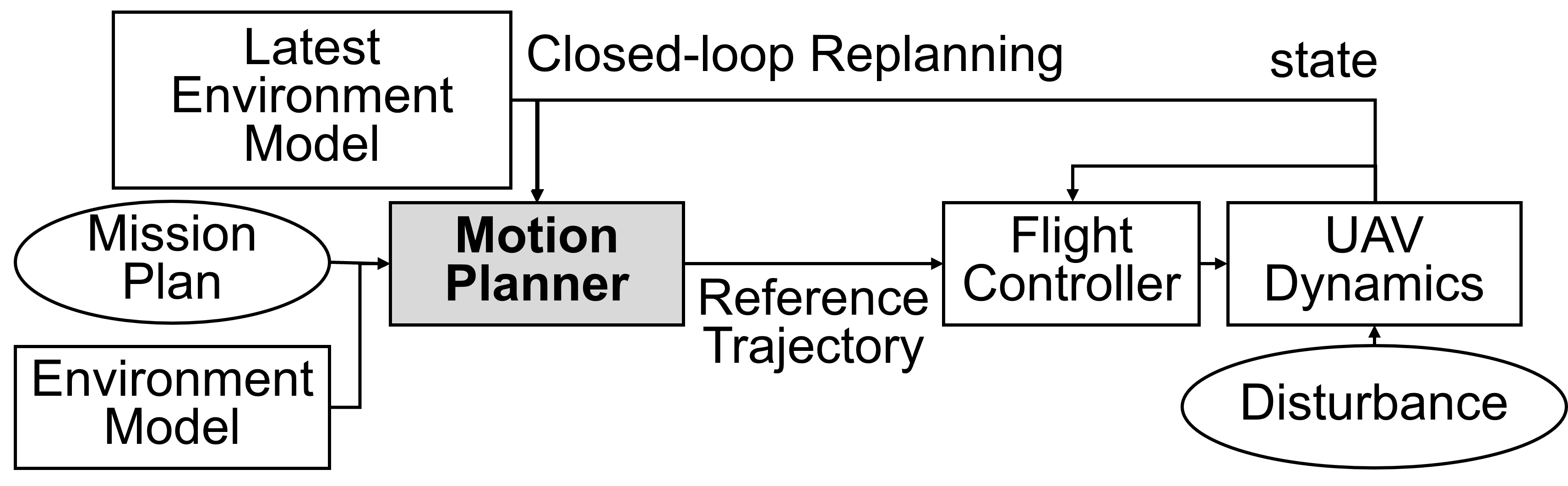}
    \caption{Block diagram representation of the trajectory planning and control framework.}
    \label{fig:controlScheme}
\end{figure}

In this section, we introduce a closed-loop re-planning framework for solving the motion planning problem formulated in the previous section. This framework unifies mission objectives (e.g., energy efficiency, path feasibility), models of the operating environment (e.g., obstacles, no-fly zones, weather), and aircraft dynamics. It provides a closed-loop architecture for generating and iteratively refining UAV trajectories. An overview is shown in Fig.~\ref{fig:controlScheme}. The open-loop planner generates a reference trajectory, which the flight controller executes to produce the corresponding control commands. The UAV’s current state and latest environment model are continuously used as feedback, allowing the planner to trigger re-planning when needed (e.g., emerging no-fly zones). This section is organized as follows: Sec.~\ref{ssec:openloop_planner} derives the nominal open-loop planner and
Sec.~\ref{ssec:closedloop_re-planning} shows the closed-loop re-planner framework. 

\subsection{Open-loop Planner}
\label{ssec:openloop_planner}
The planning algorithm is built on the open-loop planner described in this subsection. We first transform the original flight planning problem into a Markov decision process (MDP) 
 in (Sec.~\ref{sssec:MDP}) and then derive an efficient MCTS-based algorithm to solve the MDP in the open-loop setting (Sec.~\ref{sssec:MCTS}).

\subsubsection{MDP Reformulation}
\label{sssec:MDP}
To reformulate the trajectory planning problem from Eq.~\ref{eq:constrained} as a finite-horizon MDP, we define a new state variable $S \in \mathcal{S} \subset \left\{(X,v) \in \mathcal{X} \times \reals^3|p(X)\notin\Omega_n\right\}$ that augments the original state with a virtual spatial reference point $v \in \mathbb{R}^3$, that will be used to simplify the handling of navigation field (NF) constraints (discussed later in this section). We also introduce $r \in \R^3$ as the waypoint variable to be tracked, which serves as the output of the planner and the input to the flight controller. The resulting optimization objective is to find a feasible discrete state trajectory $\{S^*_i\}_{i=1}^N = (S_1^*,\dots,S_N^*)$ and a corresponding sequence of reference waypoints $\{r_i^*\}_{i=1}^N = (r_1^*,\dots,r_N^*)$ that solves the following optimization problem:

\begin{subequations} \label{eq:MDP}
\begin{align}
\max_{\{S_i\}_{i=1}^N,\,\{r_i\}_{i=1}^N} \quad & \mathcal{D}(S_N) \\[3pt]
\text{s.t.} \quad
& S_{i+1} = F_{\text{MDP}}(S_i, r_i, w), \quad i = 0,\dots,N-1, \label{eq:MDP-b} \\[3pt] 
& S_0 = (\xi, v_0), \\[3pt]
& r_i \in \R^3,\quad S_i \in \mathcal{S}
\end{align}
\end{subequations}

\noindent where $\mathcal{D}:\mathcal{S}\to\mathbb{R}_{\ge0}$ is the terminal reward, $F_{\text{MDP}}$ is a discretization of the closed-loop aircraft dynamics, and $v_0 \in \mathbb{R}^3$ denotes the initial applied reference. Different from the continuous-time variables $X(t)$ and $U(t)$, the MDP states $S_i$ and inputs $r_i$ represent the discrete-time evolution of the aircraft under closed-loop feedback.

\paragraph{Dynamics:}
\label{para:dynamics}

The dynamics in Eq.~\eqref{eq:MDP-b} are obtained by adding pre-stabilizing controllers and a navigation field (which gives the system local constraint-avoidance capabilities) to the open-loop dynamics in Eq.~\eqref{eq:dynamics}. This allows us to sample reference commands (rather than open-loop control inputs), which has been shown to greatly improve the efficiency of sampling-based algorithms \citep{riviere2021spectral}. The key idea is to use the virtual reference point $v$ that the aircraft is commanded to track. The vehicle’s spatial position is driven toward $v$ by pre-stabilizing low-level controllers. Under this formulation, the planner no longer directly selects the open-loop dynamics inputs $U = [v_a\ \gamma\ \omega]^\top$; instead, these inputs are produced by the tracking controllers described by Eq.~\ref{eq:controlMapping} later in this section.

\begin{figure}[h!]
    \centering    
    \includegraphics[width=0.5\textwidth]{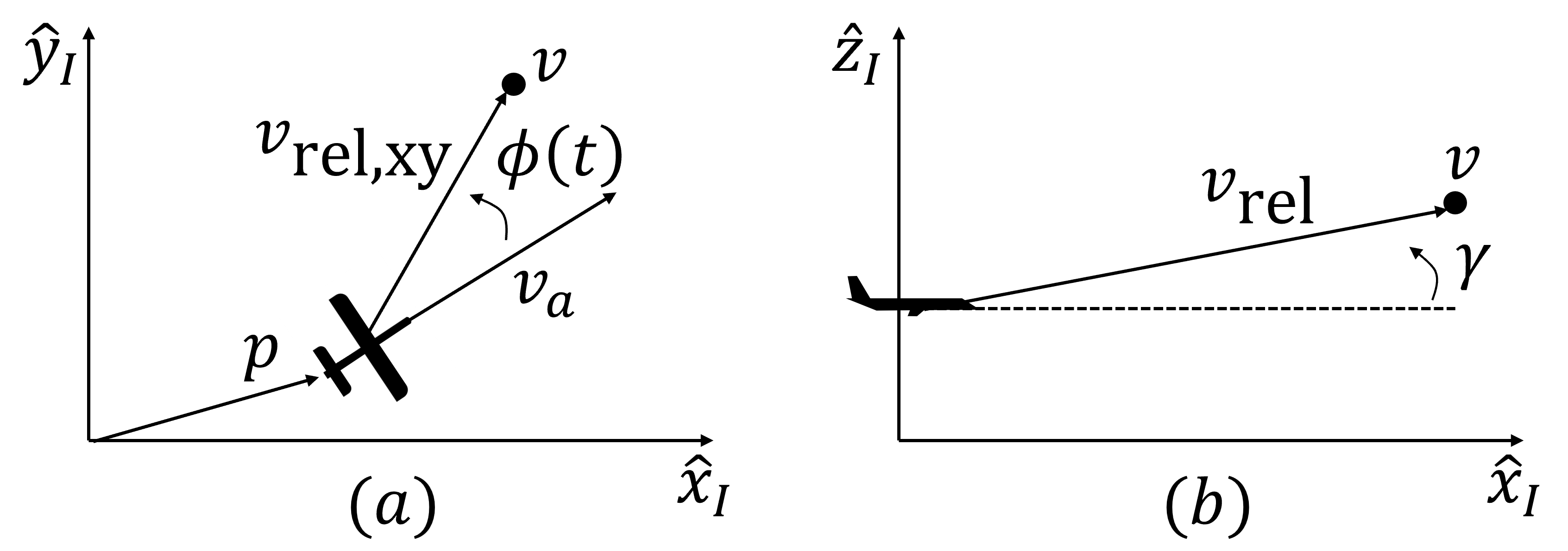}
    \caption{Graphical representation of the navigation vector from (a) a top-down view and (b) a side view of the aircraft. The inertial reference frame axes are shown as $\hat{x}_I$, $\hat{y}_I$, $\hat{z}_I$
    .}
    \label{fig:navigation}
\end{figure}

We first pre-stabilize Eqs.~\eqref{eq:kin_x}–\eqref{eq:kin_psi} to ensure tracking of the desired reference point. We regulate the heading error $\phi(t)$, the flight-path angle $\gamma(t)$, and the airspeed $v_a(t)$ to align the aircraft velocity vector with the relative navigation vector. The navigation vector $v_{\text{rel}}\in \reals ^3$ is defined as the difference between the reference point $v$ and the current aircraft position ${p}$, $v_{\text{rel}} = v - p$, as illustrated in Fig.~\ref{fig:navigation}.

First, we simply set the airspeed $v_a(t)$ to be a constant $v_a$ for the duration of the mission. Next, we chose to regulate the heading error $\phi(t)$ 

\change{
\begin{equation}
\label{eq:phi}
\phi(t)
= \cos^{-1}\!\left(
  \frac{(v-p)^\top
  [v_a\cos\psi\quad v_a\sin\psi\quad0]^\top}
       {\|(v-p)_{xy}\|\,v_a}
\right)
\end{equation}
}

\noindent through the yaw rate $\omega(t)$ using the unicycle-based control strategy \citep{lee2000stable}:

\begin{equation} \label{eq:omegaLawSimple}
    \omega(t) = -K_1 \sin\!\big(\phi(t)\big)\cos\!\big(\phi(t)\big) - K_2\, \phi(t)
\end{equation}
\noindent where $K_2$ is a proportional gain and $K_1$ weights a nonlinear stabilizing term. \change{This unicycle-based heading law uses well-established techniques from nonlinear control and is natural fit here, as it reliably stabilizes the heading error for this kinematic model with very low computational overhead.} Finally, the flight path angle $\gamma(t)$ is regulated as the signed angle between the inertial $xy$ plane and the relative navigation vector $v_{\text{rel}}$. The analytical expression for $\gamma(t)$ is given by:

\change{
\begin{equation}\label{eq:flightPath}
\gamma(t)=\mathrm{atan2}([0~0~1]v_{rel},\| Mv_{rel}\|)
\end{equation}
}

With all control inputs defined, the control law can be expressed as a function of the reference point $v$ and aircraft state $X$:

\begin{equation} \label{eq:controlMapping}  
    U = g(X,v) =[    v_a~\gamma(t)~\omega(t)
]^\top
\end{equation}

Substituting \eqref{eq:controlMapping} into the open-loop dynamics~\eqref{eq:dynamics} yields the pre-stabilized closed-loop model

\begin{equation}
\label{eq:dynamicsWithControl}
\dot{X} = f\big(X, g(X,v), w\big)
\end{equation}

\noindent which ensures stable tracking of the reference point $v$. \change{This pre-stabilization allows MCTS to sample only reference points $v$, while the low-level controller $g(X,v)$ automatically generates admissible inputs. As a result, the tree search operates in a lower-dimensional, well-constrained action space, which greatly improves sampling efficiency.}

\paragraph{Navigation Field:}
\label{para:NF}

Given the pre-stabilized UAV dynamics in Eq.~\ref{eq:dynamicsWithControl}, the next objective is to incorporate some local constraint-handling capabilities. This is achieved using a potential-based navigation field to drive the \emph{applied reference} $v(t)$ toward the \emph{desired reference} command $r(t) \in \mathbb{R}^3$ in a manner that avoids constraint violations. The navigation field functions as a local and approximate path planner and enforces constraint satisfaction by regulating the evolution of $v(t)$ according to

\begin{equation} \label{eq:referenceDynamics}
    \dot{v} = \rho_{NF}(v,r)\Delta_s
\end{equation}

\noindent where $\rho_{NF}(v, r)$ is the navigation field, which generates a collision-free path from $v$ to $r$, and $\Delta_s>0$ is a static safety margin that limits the rate at which $v$ may approach $r$. 

We use potential field theory to build the navigation field $\rho_{NF}(v,r) = \rho_{att}(v,r) + \rho_{rep}(v)$ which consists of an attraction field $\rho_{att}$ that drives $v$ toward $r$ and a repulsion field $\rho_{rep}$ that manoeuvres it around obstacles~\citep{nicotra2018explicit,convens2021invariant}. The attractive component directs the applied reference toward the desired reference according to

\begin{equation} \label{eq:NFAtt}
    \rho_{att}(v,r)=\!\left[\frac{r-v}{\max(\|r-v\|,\eta_\rho)}\right]^\top
\end{equation}

\noindent where $\eta_\rho>0$ is a small smoothing parameter that prevents numerical instability when $r\approx v$. The repulsion field pushes the reference away from nearby constraints and is obtained by summing the contributions from all constraints:

\begin{equation}
\label{eq:NFRep}
    \rho_{\text{rep}}(v) =
    \sum_{i=1}^{N_I}
    \max\!\left(\frac{\zeta - c_i(v)}{\zeta - \delta},\, 0\right)
    \frac{\nabla_v c_i(v)}{\bigl\|\nabla_v c_i(v)\bigr\|}
\end{equation}

\noindent where $\delta>0$ is the minimum allowable clearance, $\zeta>\delta$ is the influence margin, and

\begin{equation}
\label{eq:c}
\begin{aligned}
    c_1(v) &= c_{\text{terrain}}(v) = z_v - \operatorname{f}(x_v, y_v) - \varepsilon_z,\\
    c_2(v) &= c_{\text{ceiling}}(v) = z_{\max} - z_v,\\
    c_{i+2}(v) &= c_{\text{obs},i}(v)
        = \bigl\|[x_v,\,y_v]^\top - p_{c_i}\bigr\| - R_{c_i},
  \qquad i = 1,\dots,N_c
\end{aligned}
\end{equation}

\noindent are level-set representations of the terrain, ceiling, and obstacle-avoidance constraints.

By combining the closed-loop UAV dynamics~\eqref{eq:dynamicsWithControl} and the navigation-field-guided reference evolution~\eqref{eq:referenceDynamics}, the system used in the MDP formulation is

\begin{equation} \label{eq:MDPdyn}
\tilde{F}(S,w,r)=\begin{bmatrix}
\dot{X}\\ \dot{v}
\end{bmatrix}=\begin{bmatrix}
\mathcal{F}(X, g(X,v),w)\\ \rho_{NF}(v,r)\Delta_s
\end{bmatrix}
\end{equation}

This constraint-aware formulation ensures that the applied reference point $v(t)$ evolves so that, as the UAV tracks $v(t)$ toward $r$, the state $X(t)$ approximately remains within the feasible region. During planning, the NF is embedded in the dynamics used in all trajectory simulations, such as MCTS tree expansion and simulation, ensuring that most of the sampled trajectories are dynamically feasible and compliant with operational constraints.

\paragraph{MDP Transition Function:}
\label{para:F_MDP}

The transition function $F_{\text{MDP}}$ represents the propagation of the closed-loop dynamics \eqref{eq:MDPdyn} from $S_i$ until the aircraft state catches the applied reference associated with the next desired waypoint $r_i$.

\begin{subequations} \label{eq:MDP_transition}
\begin{align}
S_{i+1} &= F_{\text{MDP}}(S_i, r_i, w) = S(\hat{T},S_i,r_i,w), \\[3pt]
S(t,S_0,r,w) &= S_0 + \int_{0}^{t} \tilde{F}(S(\tau), w, r)\,d\tau, \\[3pt]
\hat{T} &= \inf_t\left\{\,t \geq 0 ~\big|~ \Big\|h\big(S(t,S_i,r_i,w)\big) - h_v\big(S(t,S_i,r_i,w)\big)\Big\| \leq \varepsilon \right\}
\end{align}
\end{subequations}

\noindent where $h(S)$ extracts the spatial position of the aircraft and $h_v(S)$ extracts the applied reference position from the augmented state.

\paragraph{Reward Function:}
\label{para:reward}

The terminal reward to be maximized is the remaining onboard energy when the UAV reaches the goal region, and zero if the trajectory becomes infeasible. Starting from $S_N$, the UAV state at the end of the planning horizon, the controllers drive the system to the goal region, yielding a final state $S_{\text{f}} = F_{\text{MDP}}(S_N,p_t,w)$ from which we can extract the energy $E_{\text{f}} = [0~0~0~0~1~0~0~0]^T S_{\text{f}}$, so that

\begin{equation} \label{eq:terminalReward}
    \mathcal{D}(S_{N}) = \begin{cases} 
    [0~0~0~0~1~0~0~0]^T S_{\text{f}} & \text{if}\quad h(S_{\text{f}})\in P_T\\
    0 & \text{else}
    \end{cases}
\end{equation}

\subsubsection{MCTS Algorithm}
\label{sssec:MCTS}

The objective of the MCTS planner is to find an approximate solution to the MDP~\eqref{eq:MDP} by iteratively simulating and evaluating candidate waypoint sequences. MCTS uses a tree structure to represent possible flight paths~\citep{lathrop2023mpt,riviere2021spectral}. Each node corresponds to a UAV state at a given time, while edges represent feasible transitions between waypoints governed by the aircraft’s closed-loop dynamics. The depth of the MCTS tree is the planning horizon $N$, and each layer represents one additional waypoint. Each node $i$ stores the discrete state $S_i \in \mathcal{S}$, a cumulative reward estimate $Q_i \in \R$, a visit count $N_i \in \mathbb{N}$, a parent index $parent(i)$, and its set of child nodes $children(i)$.

\begin{figure}[h!]
    \centering
    \includegraphics[width=0.5\textwidth]{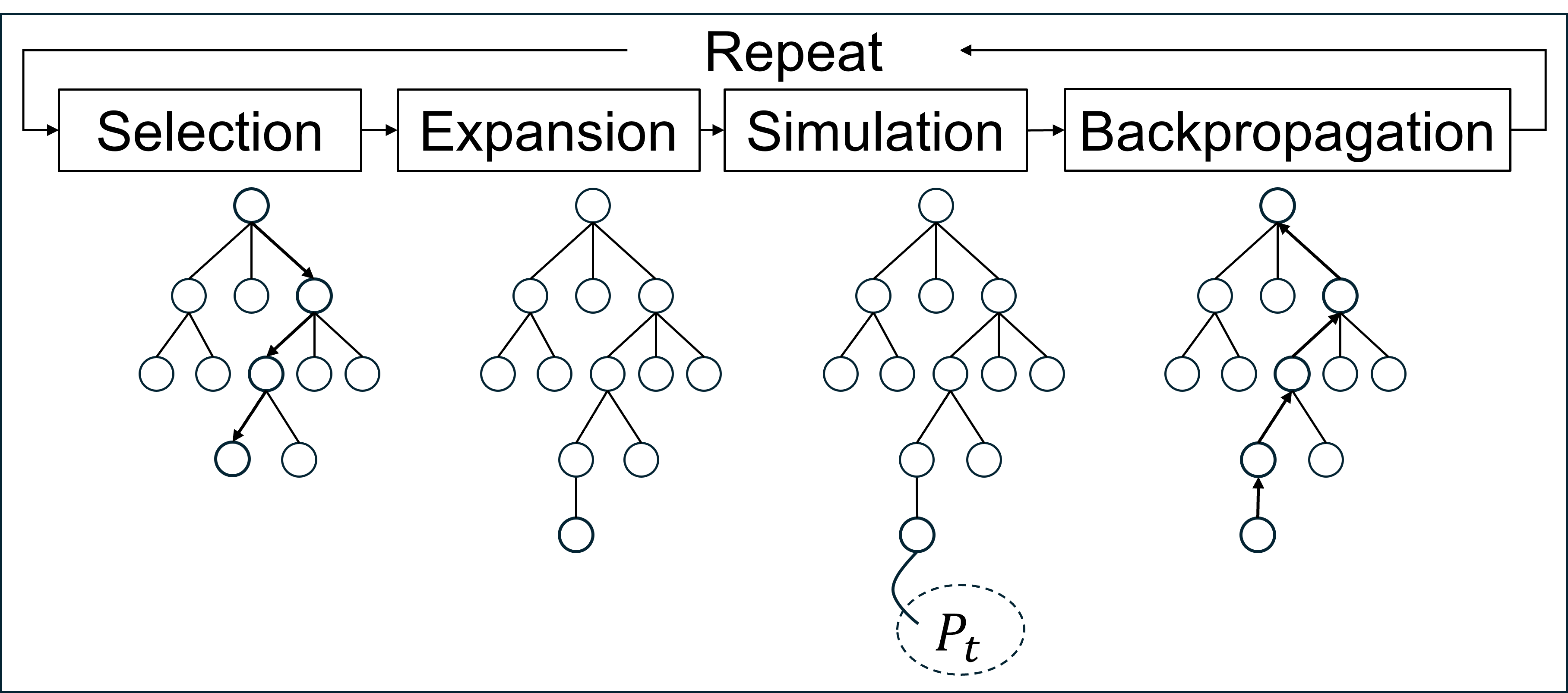}
    \caption{Flowchart of the proposed MCTS planner procedure.}
    \label{fig:MCTS_planner_flowchart}
\end{figure}

The MCTS planner proceeds iteratively through four main phases: \textit{Selection}, \textit{Expansion}, \textit{Simulation}, and \textit{Backpropagation}, as illustrated in Fig.~\ref{fig:MCTS_planner_flowchart}. Each phase is described by the following: 

\begin{enumerate}

    \item \textbf{Selection}: Starting from the root, the algorithm recursively selects child nodes based on the Upper Confidence bounds applied for Trees (UCT) policy (Line 15 in Algorithm~\ref{alg:MCTSPlanner}), where $c_\text{UCT}$ is the coefficient that balances exploration and exploitation. This mechanism ensures that promising branches are explored while maintaining sufficient search diversity~\citep{shah2020nonasymptotic, silver2017mastering}.

    \item \textbf{Expansion}: For each selected child node $i$, a discrete set of candidate reference points is precomputed by uniform sampling of headings and fixed radial distance subject to the aircraft’s flight envelope constraints. Specifically, we define sets of horizontal and vertical angle offsets $\Delta\Psi = \{-\Delta\psi_{\max},\dots,0,\dots,\Delta\psi_{\max}\}$, $\Delta\Gamma = \{-\Delta\gamma_{\max},\dots,0,\dots,\Delta\gamma_{\max}\}$, and a fixed look-ahead distance $d_{\mathrm{ref}} > 0$. For each pair $(\Delta\psi_k,\Delta\gamma_\ell)\in\Delta\Psi\times\Delta\Gamma$, we construct a candidate reference point 

    \begin{equation} \label{eq:generateRefs}
    r_{k,\ell} = h(S_i) +
    d_{\mathrm{ref}}
    \begin{bmatrix}
    \cos(\gamma_i + \Delta\gamma_\ell)\cos(\psi_i + \Delta\psi_k) \\
    \cos(\gamma_i + \Delta\gamma_\ell)\sin(\psi_i + \Delta\psi_k) \\
    \sin(\gamma_i + \Delta\gamma_\ell)
    \end{bmatrix}
    \end{equation}
    
    \noindent where $\gamma_i$ and $\psi_i$ are the flight-path and heading states of node $i$. The indices $k$ and $\ell$ enumerate the horizontal and vertical angle offsets, and the angles are chosen so that the corresponding commands remain within $\mathcal{U}$ in~\eqref{eq:inputConstraints} when propagated without wind or obstacles.

    A selected node with at least one unexplored candidate reference point and with depth less than $N$ is expanded by generating a new state $S_{\text{new}}$ via the closed-loop dynamics~\eqref{eq:MDPdyn} from the current node state $S_i$ toward a newly selected reference point $r_s$ (Line 9 in Algorithm~\ref{alg:MCTSPlanner}). A node is considered fully expanded once all candidate reference points have been explored, and only transitions that yield feasible states are retained in the tree.

    \item \textbf{Simulation}: When a path segment is completed (i.e., when the tree depth reaches $N$, yielding a path of $N+1$ nodes), a forward simulation is performed from the leaf-node state toward the mission target $p_t$ to evaluate the terminal reward~\eqref{eq:terminalReward}.

    \item \textbf{Backpropagation}: The obtained reward is propagated upward through the tree to update each ancestor node’s cumulative reward $Q_i$ and visit count $N_i$. This iterative update refines the value estimates for each trajectory branch and guides subsequent search iterations toward more energy-efficient trajectories.

\end{enumerate}

Through repeated iterations, the planner incrementally refines its search tree, which naturally grows denser in high-reward regions, converging toward a \change{near-optimal} reference policy that minimizes total energy consumption while satisfying environmental and kinematic constraints. The detailed implementation of the MCTS procedure is presented in Algorithm~\ref{alg:MCTSPlanner}. In Line 3, the search tree "$tree$" is initialized with the root node corresponding to the initial state of the UAV. The MCTS function then iteratively performs the four MCTS phases until the CPU budget is exhausted. In Line 5, a new search path is initialized with the root node. Lines 6–17 construct the path through the \textit{Selection} and \textit{Expansion} phases. The feasibility of $S_{\text{new}}$ is then checked, and if valid, a new child node is added. In Lines 15–16, the child node with the highest UCT value is selected and appended to the path. Once the path is complete, Line 18 performs the \textit{Simulation} phase to compute the terminal reward based on the terminal state. Finally, Lines 19–21 execute the $\textit{Backpropagation}$ phase, updating the visit counts and cumulative rewards of all nodes along the path.

\begin{algorithm}
\caption{MCTS Trajectory Planning}
\label{alg:MCTSPlanner}
\begin{algorithmic}[1]
\State initialize root node state $S_0$ with initial state $\xi$
\Function{MCTS}{$S_0$, $\Omega_n$, $w$}
    \State initialize $tree$ with root node state $S_0$ and $\text{best\_path} \gets [\,]$
    \While{not terminated}
        \State $path \gets [tree.\text{root}]$, $i \gets tree.\text{root}$
        \While{path not completed}
            \If{$i$ not expanded}
                \State Sample random reference $r_s$ from $i$
                \State $S_{\text{new}} \gets F_{\text{MDP}}(S_i,~r_s,~w)$
                \If{$S_{\text{new}}$ is not feasible}
                    \State \textbf{break}
                \EndIf
                \State Create new node with state $S_{\text{new}}$ and add to $\text{children}(i)$
            \EndIf
            \State $i \gets \arg\max\limits_{i^* \in children(i)} 
            \left( 
            \frac{Q_{i^*}}{N_{i^*}} + c_\text{UCT} \sqrt{\frac{N_i}{N_{i^*}}}
            \right)$
            \Comment{Choose the best child according to UCT}
            \State Append $i$ to path
        \EndWhile
        \State $reward \gets E\big(F_{\text{MDP}}(S_i, p_t, w)\big)$
        \Comment{Simulation from the leaf node to mission target}
        \For{each node $i$ in path}
            \State $N_j \gets N_j + 1$, \quad $Q_j \gets Q_j + reward$
        \EndFor
        \State Update $\text{best\_path}$ if $reward$ improves
    \EndWhile
    \State \Return $\text{best\_path}$ and $h(\text{best\_path})$
\EndFunction
\end{algorithmic}
\end{algorithm}

\paragraph{Initial Flight Plan:}
\label{para:olp}

The planner first generates a complete flight plan prior to takeoff, providing a sequence of waypoint commands $\{r_i\}_{i=1}^N$ as the nominal mission plan for the UAV.

The planner constructs an MCTS search tree incorporating the latest available environmental data (e.g., $w_{t_0}$ for the wind forecast and $\Omega_{n,t_0}$ for the no-fly-zone forecast) and executes the MCTS procedure described in Sec.~\ref{sssec:MCTS}. Because the initial flight plan is run offline prior to takeoff, we can run the MCTS algorithm until convergence.

\begin{figure}[h!]
    \centering
    \includegraphics[width=0.5\textwidth]{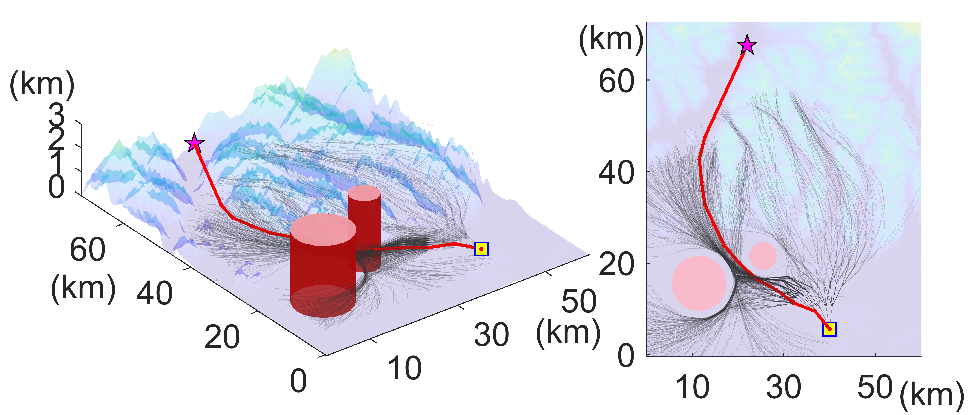}
    \caption{Example of a flight plan generated by the MCTS planner. Gray branches indicate explored paths, and the red path denotes the selected \change{near-optimal} trajectory.}
    \label{fig:openloop_path_example}
\end{figure}

An example output of the open-loop MCTS planner is shown in Fig.~\ref{fig:openloop_path_example}, where the algorithm successfully identifies a smooth and feasible trajectory that balances terrain clearance, wind exploitation, and energy efficiency.

\subsection{Closed-loop Re-planning}
\label{ssec:closedloop_re-planning}

Although the open-loop solution is \change{near-optimal} with respect to the pre-flight forecast, deviations may arise during flight due to dynamic environmental changes, motivating the need for closed-loop correction strategies (i.e, re-planning). The UAV’s low-level flight controllers can compensate for small disturbances such as minor wind fluctuations, but significant environmental changes (e.g., storms, large-scale wind shifts, or intense precipitation) might make the initial flight plan infeasible. The proposed MCTS planner can be naturally extended into a closed-loop re-planner by repeatedly generating flight plans over a receding horizon as is done in model predictive control. At each re-planning step, MCTS generates a sequence of waypoints based on the current aircraft state and forecast, the UAV tracks only the first waypoint, then the planning problem is resolved using updated state and forecast information, and the process repeats.

Let $\left(\{S_i^*\}_{i=1}^N,~\{r_i^*\}_{i=1}^N\right) = \mathrm{MCTS}(S_0,~\Omega_{n,t},~w_t)$ denote the \change{near-optimal} discrete state trajectory and associated reference
waypoints obtained by solving the MDP~\eqref{eq:MDP} with the MCTS planner (Algorithm~\ref{alg:MCTSPlanner}), where $\Omega_{n,t}$ and $w_t$ denote the no-fly-zone and wind forecasts at time $t$. The planner outputs a sequence of waypoints $\{r_i^*\}_{i=1}^N$ and sequence of predicted states $\{S_i^*\}_{i=1}^N$. At each re-planning step, the first waypoint $r_1^*$ is passed to the onboard closed-loop controller, which tracks this waypoint using the pre-stabilized dynamics~\eqref{eq:dynamicsWithControl}. Re-planning is triggered once the current waypoint is reached (within a specified tolerance), ensuring that the executed trajectory continuously adapts to evolving environmental conditions.

The online re-planning procedure is summarized in Algorithm~\ref{alg:clp}. The algorithm takes the initial vehicle state and environmental forecasts as inputs, generates an initial flight plan, and selects the first waypoint as the active tracking target when the mission starts (Lines~1–5). During execution, the UAV tracks the current waypoint $r_{\text{wp}}$ while its state evolves until the waypoint is reached within tolerance or the maximum re-planning interval elapses (Lines~7–10). In parallel, the MCTS planner keeps running in the background refining the flight plan from the current subtree rooted at $S_1^*$ (Line~11). At each re-planning epoch, the current state and updated forecasts are incorporated (Line~12). If the vehicle deviates beyond the reuse tolerance or the forecasts change significantly, the existing tree is deemed unreliable, the root is reinitialized at the current state, and a new plan is computed (Lines~13–17). The first waypoint of the updated plan is then set as the new tracking target, and this loop continues until the terminal region is reached.

\begin{algorithm}
\caption{Closed-loop MCTS Re-planning}
\label{alg:clp}
\begin{algorithmic}[1]
\State \textbf{Input:} Initial state $\xi$; initial forecasts $w_{t_0}$, $\Omega_{n,t_0}$
\State \textbf{Parameters:} Waypoint tolerance $\varepsilon_{\text{wp}}$; maximum re-planning interval $t_{max}$; tree-reuse tolerance $\varepsilon_{\text{reuse}}$
\State Initialize root node state $S_0$ with $\xi$
\State $(\{S_i^*\}_{i=1}^N, \{r_i^*\}_{i=1}^N) \gets \mathrm{MCTS}(S_0,\,\Omega_{n,t_0},\,w_{t_0})$ 
\Comment{Initial flight plan}
\State Set current waypoint $r_{\text{wp}} \gets r_1^*$ and start mission
\While{$X(t) \notin P_T$}
    \While{$\|h(X(t)) - r_{\text{wp}}\| > \varepsilon_{\text{wp}}$ \textbf{and} $t - t_0 < t_{max}$}
        \State $U \gets g\big(X(t), r_{\text{wp}}\big)$
        \State $\dot{X} \gets \mathcal{F}(X, U, w)$
    \EndWhile
    \State $(\{S_i^*\}_{i=1}^N, \{r_i^*\}_{i=1}^N) \gets \mathrm{MCTS}(S_1^*,\,\Omega_{n,t},\,w_t)$
    \Comment{Background re-planning from the current subtree}
    \State Measure current state $X(t)$ and acquire updated forecasts $w_t, \Omega_{n,t}$
    \If{$\|h\big(X(t)\big) - r_{\text{wp}}\| > \varepsilon_{\text{reuse}}$ \textbf{or} $\Omega_{n,t} \neq \Omega_{n,t-1}$ \textbf{or} $w_t \neq w_{t-1}$}
        \Comment{Tree reuse no longer reliable}
        \State Reinitialize root node state $S_0$ with $X(t)$
        \State $(\{S_i^*\}_{i=1}^N, \{r_i^*\}_{i=1}^N) \gets \mathrm{MCTS}(S_0,\,\Omega_{n,t},\,w_t)$
        \Comment{Regenerate plan from the current state}
    \EndIf
    \State Update $r_{\text{wp}} \gets r_1^*$
    \State Reset $t_0 \gets t$
\EndWhile
\end{algorithmic}
\end{algorithm}

\paragraph{Reusing Trees}
\label{para:tree_reuse}

\begin{figure}[h!]
    \centering
    \includegraphics[width=0.5\textwidth]{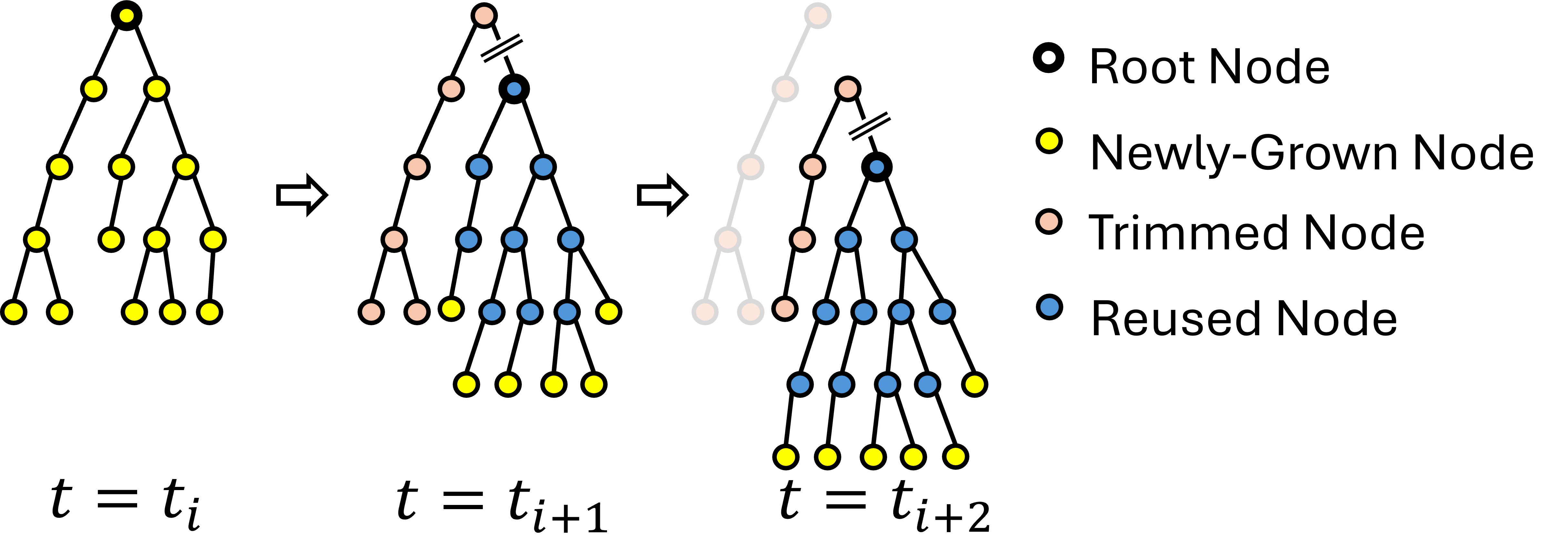}
    \caption{Illustration of tree reuse during closed-loop re-planning.}
    \label{fig:treeReuse}
\end{figure}

Solving the MDP~\eqref{eq:MDP} repeatedly in real time can be computationally demanding. To mitigate this, the re-planner exploits the structural similarity between consecutive planning problems by reusing parts of the search tree. Inspired by Model Predictive Trees (MPT)~\citep{lathrop2023mpt}, subtree reuse is enabled whenever the disturbances are small enough for the low-level controllers to correct and the environment forecasts are unchanged. In this case, the subtree rooted at the branch corresponding to the current best trajectory is retained, the MCTS root is shifted to the node $S_{1}^*$, only its sibling branches are pruned, and the informed subtree is further expanded (Lines~11 of Algorithm~\ref{alg:clp}), reducing computational overhead and preserving continuity between successive plans. The workflow is illustrated in Fig.~\ref{fig:treeReuse}, where the retained subtree (blue nodes) serves as the new root for expansion, enabling efficient online adaptation while preserving trajectory smoothness.

However, subtree reuse is not always applicable. When major environmental updates (e.g., large wind shifts or new no-fly zones) invalidate parts of the existing tree, or when the actual state deviates excessively from the desired waypoint, the re-planner discards the outdated structure and rebuilds a new tree from the current UAV state, as indicated in Lines~13–16 of Algorithm~\ref{alg:clp}. The newly generated flight plan then defines the next waypoint (Line~17), and the re-planning algorithm proceeds with this updated tree.

\section{Simulations and Results}
\label{sec:simulation_results}

To evaluate the performance of the proposed MCTS-based motion planner, we present two case studies inspired by realistic UAV deployment scenarios.

\paragraph{Implementation Details}

All simulations are performed in MATLAB 2024a on a 4.0~GHz AMD octa-core CPU. Aircraft trajectories are integrated using MATLAB’s \texttt{ode113} solver. To emulate forecast uncertainty during tracking, zero-mean Gaussian noise with variance $2~(\mathrm{m}/\mathrm{s})^2$ is added to the wind field used in the UAV dynamics simulations, while the planner operates on the corresponding nominal wind forecast. Terrain topography is obtained from the provincial digital elevation model for British Columbia~\citep{bc_dem}. We consider a fixed-wing UAV with a $3.5$~m wingspan and approximately $2.5$~h endurance, representative of current industrial designs for long-range missions (e.g., the SKYDRONES Skylane 350 or Airbus Aliaca).  Only the cruise phase of flight is simulated. Aerodynamic and environmental parameters for the power model, and unicycle heading-control gains, are summarized in Table~\ref{tab:parameters}.

\begin{table}[hbt!]
  \caption{\label{tab:parameters}Key simulation parameters and notation.}
  \centering
  \begin{tabular}{lccc}
    \hline\hline
    Parameter & Symbol & Value & Unit \\
    \hline
    air speed of the aircraft & $v_a$ & 25 & m/s \\
    aircraft mass & $m$ & 35 & kg \\
    propulsion efficiency & $\eta$ & \change{0.8} & -- \\
    baseline power draw & $P_0$ & 30 & W \\
    wing surface area & $A$ & 1.4 & m$^2$ \\
    zero-lift drag coefficient & $C_{D0}$ & 0.01 & -- \\
    induced drag factor & $K$ & 0.015 & -- \\
    sea-level air density & $\rho_0$ & 1.225 & kg/m$^3$ \\
    sea-level temperature & $T_0$ & 288.15 & K \\
    temperature lapse rate & $k$ & $-6.5\times10^{-3}$ & K/m \\
    gas constant & $R$ & 287.1 & J/(kg$\cdot$K) \\
    altitude ceiling & $z_{\max}$ & 3000 & m \\
    unicycle controller gains & $K_1, K_2$ & 5, 0.7 & \change{$\mathrm{rad}/\mathrm{s}$, $\mathrm{s}^{-1}$} \\
    \hline\hline
  \end{tabular}
\end{table}

\subsection{Case 1: Long-Horizon Planning with Time-Varying Wind} 
\label{ssec:case1}

This case study evaluates the ability of the proposed MCTS-based planner to generate energy-efficient, wind-aware trajectories over a long mission horizon in a partially dynamic environment. The scenario used for this case study simulates a medical delivery from Surrey Memorial Hospital $(49^\circ 10' 36.2''\mathrm{N},\ 122^\circ 50' 42.4''\mathrm{W})$ to Squamish General Hospital $(49^\circ 41' 51.0''\mathrm{N},\ 123^\circ 09' 28.9''\mathrm{W})$. Static no-fly zones are enforced around Vancouver International Airport with a radius of $6~\mathrm{km}$ and around Vancouver Harbour with a radius of $3~\mathrm{km}$. 

\begin{figure}[h!]
    \centering
    \includegraphics[width=0.5\textwidth]{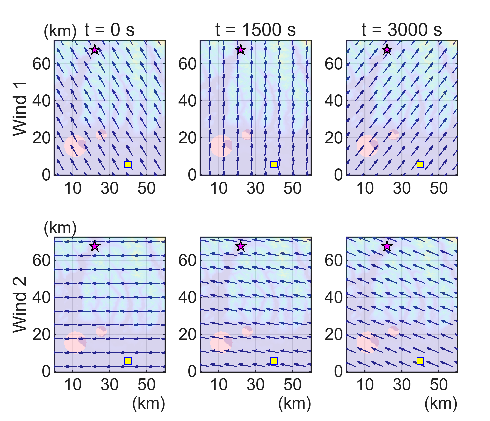}
    \caption{An aerial view of the two wind profiles used in simulation at various points in time.}
    \label{fig:windProfiles_quiver}
\end{figure}

\begin{figure}[h!]
    \centering
    \includegraphics[width=0.48\textwidth]{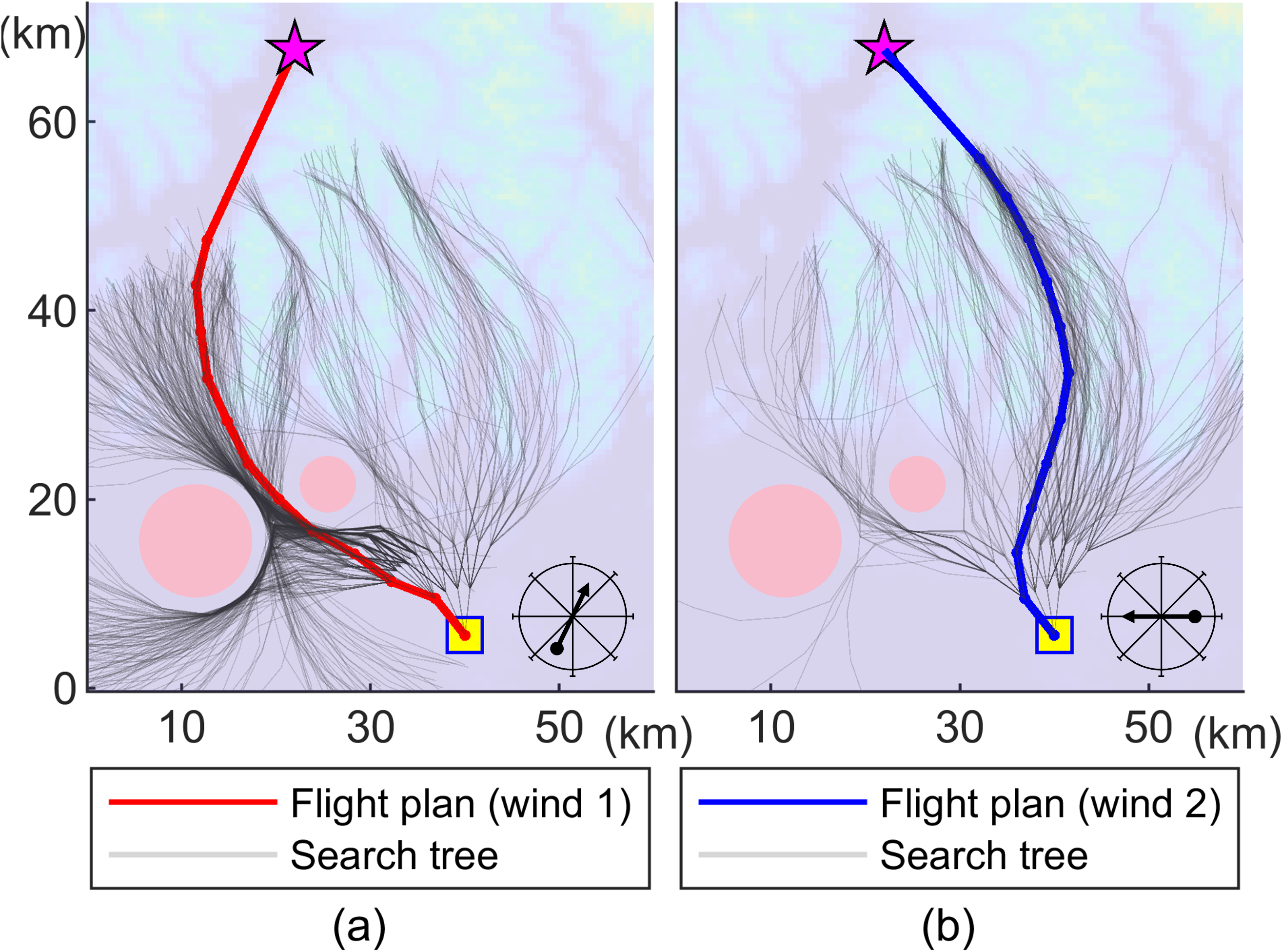}
    \caption{
    Exploration patterns of the MCTS search trees.
    (a)~Wind~1: exploration concentrates along the coastal corridor.
    (b)~Wind~2: exploration shifts toward inland valleys.
    }
    \label{fig:case1-trees}
\end{figure}

\begin{figure}[h!]
    \centering
    \includegraphics[width=0.48\textwidth]{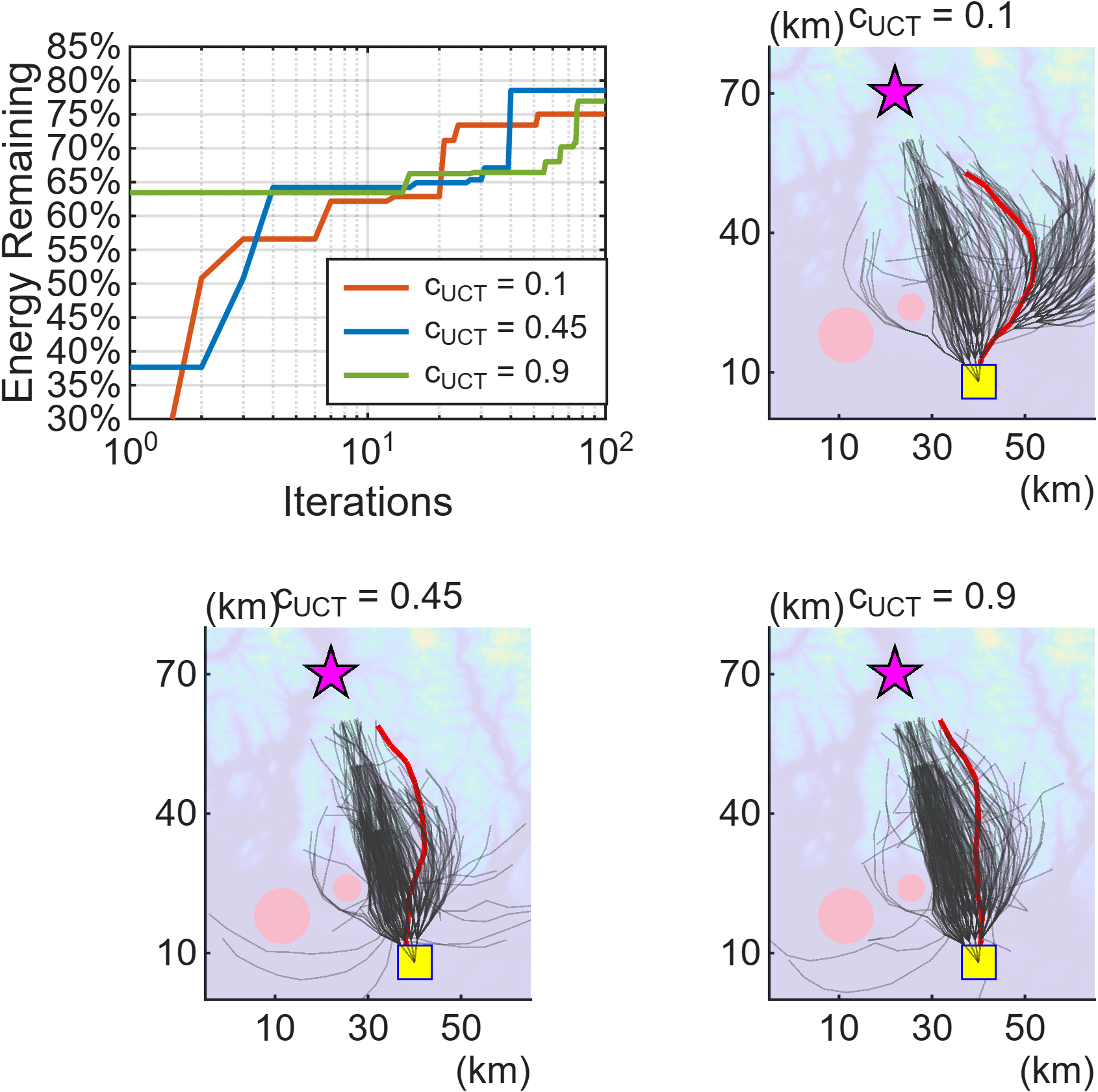}
    \caption{
    Sensitivity of MCTS to the exploration coefficient: (a) reward convergence and (b)--(d) the resulting search trees for $c_\mathrm{UCT}=0.1$, $0.45$, and $0.9$. 
    }
    \label{fig:cUCTsensitivity}
\end{figure}

\paragraph{Performance of MCTS} As shown in Fig.~\ref{fig:case1-trees}, the MCTS planner balances exploration and exploitation by sampling broadly while concentrating its search in promising corridors, and the tree expansion consistently respects the prescribed no-fly zones. The MCTS planner adapts its route based on wind evolution: under wind~1, it routes along coastal corridors to exploit favourable tailwinds, while under wind~2, crosswinds make coastal travel less beneficial and the planner selects a valley-following route with smoother elevation profiles. \change{The sensitivity of MCTS to the exploration coefficient is illustrated in Fig.~\ref{fig:cUCTsensitivity}: terminal performance is largely insensitive to $c_{\mathrm{UCT}}$ across the tested range $[0.1, 0.9]$, with all values converging to similar rewards; $c_{\mathrm{UCT}}=0.45$ was selected as it yielded marginally the best empirical convergence.}

\paragraph{Comparisons with baselines} We compare our method against \change{three baseline algorithms representative of the two dominant families of motion-planning approaches including the Batch Informed Trees (BIT*) algorithm~\citep{gammell2020bitstar}, an efficient sampling-based geometric path planner and Genetic Algorithm (GA) and Particle Swarm Optimization (PSO) methods\footnote{Both as implemented in the MATLAB 2024a Global Optimization Toolbox.}, global optimization algorithms that are widely used for fixed-wing UAV trajectory design~\citep{roberge2013GA_PSO_uav} and attempt to solve the same optimization problem \eqref{eq:constrained} as our MCTS algorithm.}

\change{All planners are stopped when the cost function converges within a tolerance of $5\times 10^{-3}$ or when the wall-clock time hits 30~minutes. The PSO and GA methods cannot dynamically add waypoints when discretizing \eqref{eq:constrained} and were provided with $N_{\mathrm{wp}}=10$ intermediate waypoints; the GA and PSO use a population (swarm) size of $40$.} The MCTS planner uses $c_{\mathrm{UCT}}=0.45$, an expansion horizon of $5000~\mathrm{m}$, and a branching factor of $12$. BIT* is configured with a step size of $1800~\mathrm{m}$ and a connection radius of $2000~\mathrm{m}$.

\begin{figure}[h!]
    \centering
    \includegraphics[width=0.48\textwidth]{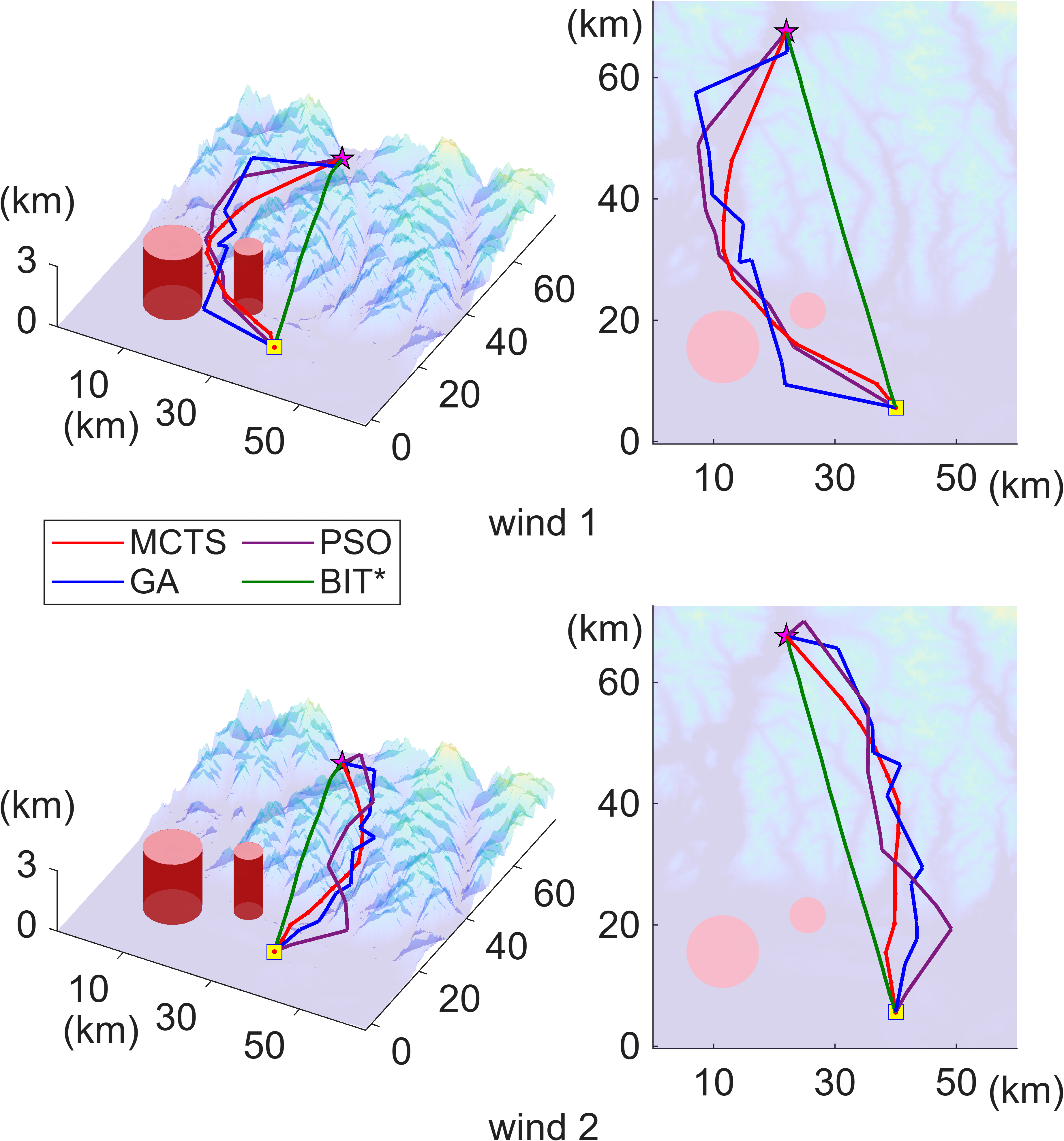}
    \caption{
       Comparison of flight plan trajectories: (a) wind~1 and (b) wind~2, for BIT* (green), GA (blue), PSO (purple), and MCTS (red). BIT* converges to a shortest-distance feasible path while MCTS, GA, and PSO algorithms converge to paths that follow valleys and exploit the prevailing wind.
    }
    \label{fig:case1-trajectories}
\end{figure}

\begin{figure}[h!]
    \centering
    \includegraphics[width=0.48\textwidth]{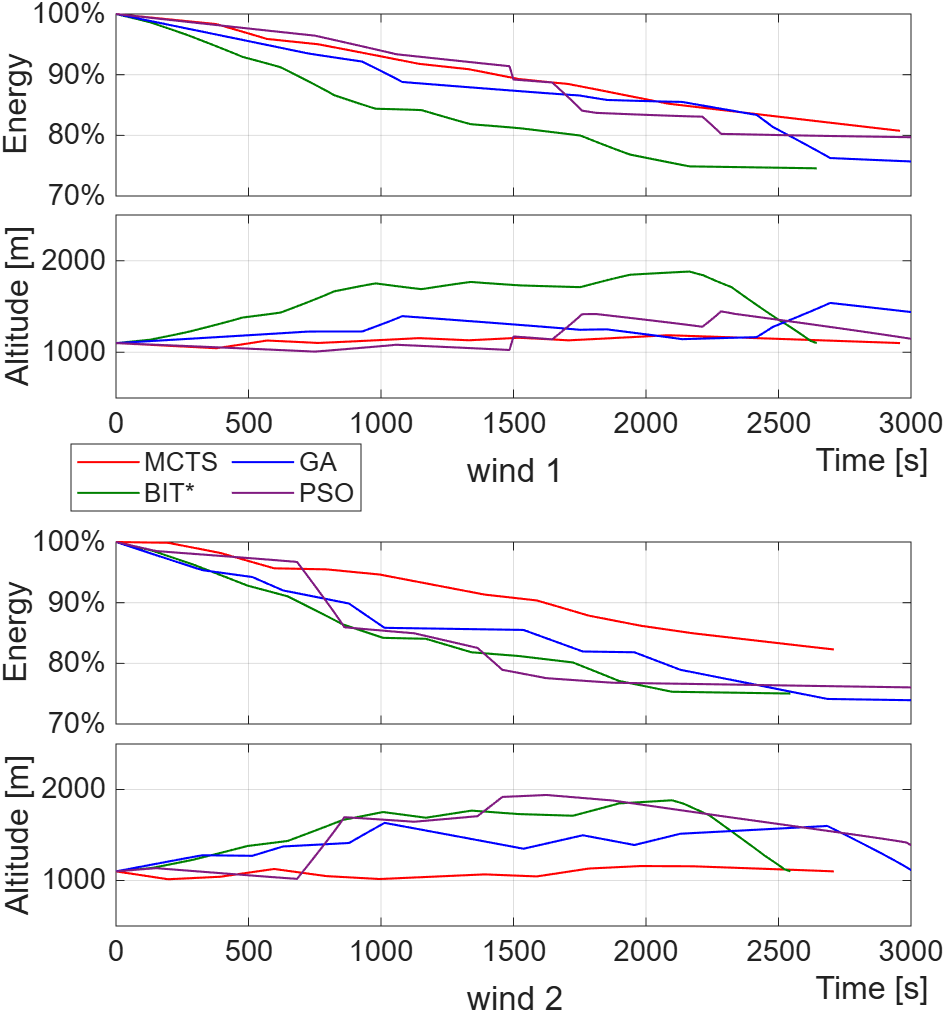}
    \caption{
    Comparison of flight plan performance for (a) wind~1 and (b) wind~2, showing onboard energy and altitude versus time.
    }
    \label{fig:case1-performance}
\end{figure}

\begin{table}[hbt!]
  \caption{\label{tab:case1_table}Quantitative results for Case~1.}
  \centering
  \begin{tabular}{lcccc}
    \hline\hline
    Method & Path length [$\mathrm{m}$] & Runtime [$\mathrm{s}$] & Energy remaining [$\%$] & Arrival time [$\mathrm{s}$] \\
    \hline
    MCTS (wind 1) & 77765 & 817.7  & 80.7 & 2958.8 \\
    MCTS (wind 2) & 68336 & 900.7  & 82.3 & 2709.1 \\
    GA (wind 1)   & 95600 & 1811.6 & 74.8 & 3006.6 \\
    GA (wind 2)   & 75993 & 1928.7 & 74.9 & 2865.3 \\
    PSO (wind 1)  & 81733 & 1802.5 & 79.6 & 3097.7 \\
    PSO (wind 2)  & 80339 & 1802.5 & 75.9 & 3172.9 \\
    BIT* (wind 1) & 64612 & 102.2 & 74.6 & 2644.4 \\
    BIT* (wind 2) & 64656 & 102.2 & 75.0 & 2545.0 \\
    \hline\hline
  \end{tabular}
\end{table}

The resulting flight plans and performance metrics for Case 1 are summarized in Fig.~\ref{fig:case1-trajectories}-\ref{fig:case1-performance} and Table~\ref{tab:case1_table}. The BIT* solution follows a near-geodesic line across mountainous terrain, ignoring wind conditions and extra energy cost of climbing. \change{The MCTS, GA, and PSO algorithms produce qualitatively similar solutions, that avoid climbing and exploit prevailing winds, however the GA and PSO algorithms are significantly less computationally efficient than the MCTS algorithm and are unable to fully converge within the provided 30 minutes\footnote{As a sanity check, we initialized the GA and PSO from the MCTS solution, and both converged to something similar to the original MCTS solution.}.} Although MCTS trajectories are geometrically longer (by $20.4\%$ and $5.7\%$ for wind~1 and wind~2 respectively relative to BIT*), they achieve the highest remaining energy upon arrival. Specifically, MCTS retains $80.7\%/82.3\%$ of onboard energy, compared with $79.6\%/75.9\%$ for PSO, $74.8\%/74.9\%$ for GA, and $74.6\%/75.0\%$ for BIT* under wind~1/wind~2, respectively. \change{This advantage comes mostly from reduced drag since favourable winds raise ground speed and shorten time aloft, lowering integrated drag and baseline power}, however the climb term in Eq.~\eqref{eq:power} dissipates a non-negligible amount of energy as the non-regenerative model prevents recovering potential energy on descent, slightly favouring valley following routes.

\begin{figure}[h!]
    \centering
    \includegraphics[width=0.48\textwidth]{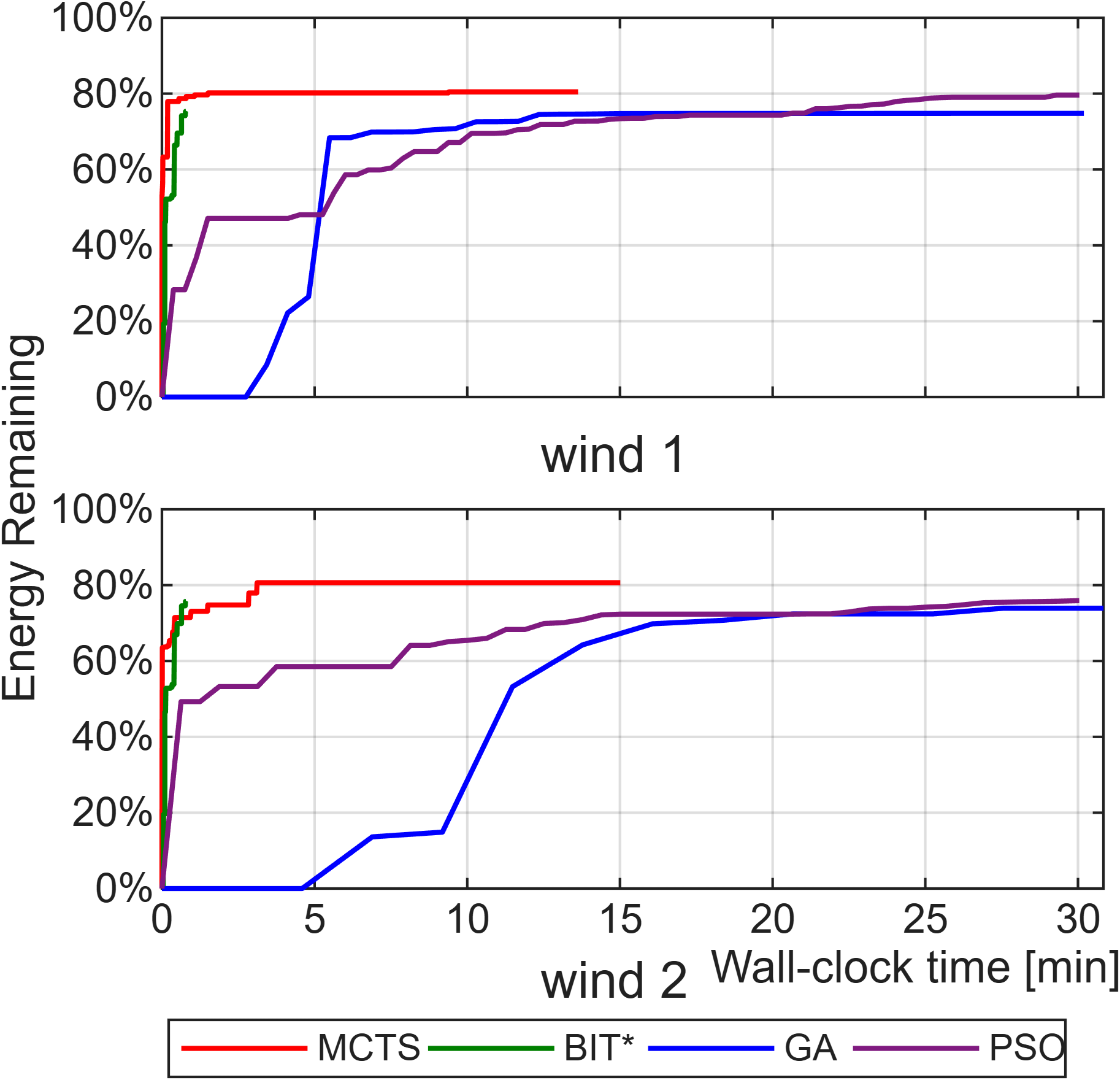}
    \caption{ 
    Reward convergence versus wall-clock time for (a) wind~1 and (b) wind~2, with BIT* converging the fastest, followed by MCTS. GA and PSO reach the 30-minute limit without converging and plateau below MCTS under both wind conditions.
    }
\label{fig:case1_convergence}
\end{figure}

\change{Figure~\ref{fig:case1_convergence} shows clear differences in how quickly the methods converge. BIT* converges fastest, since its search space shrinks progressively toward the shortest geometric path and each iteration is inexpensive. MCTS also converges quickly, reaching near-optimal reward within a small fraction of the total computation time; later iterations yielded diminishing returns. The GA and PSO, by contrast, converge relatively slowly, unlike MCTS they do not accumulate value estimates across evaluations, so they cannot concentrate search effort on promising routes. These differences are reflected in the runtimes reported in Table~\ref{tab:case1_table}: BIT* and MCTS terminate upon convergence within minutes, whereas the GA and PSO exhaust the full 30-minute budget without satisfying the convergence criterion and plateau below MCTS under both wind conditions.}

\begin{figure}[h!]
    \centering
    \includegraphics[width=0.48\textwidth]{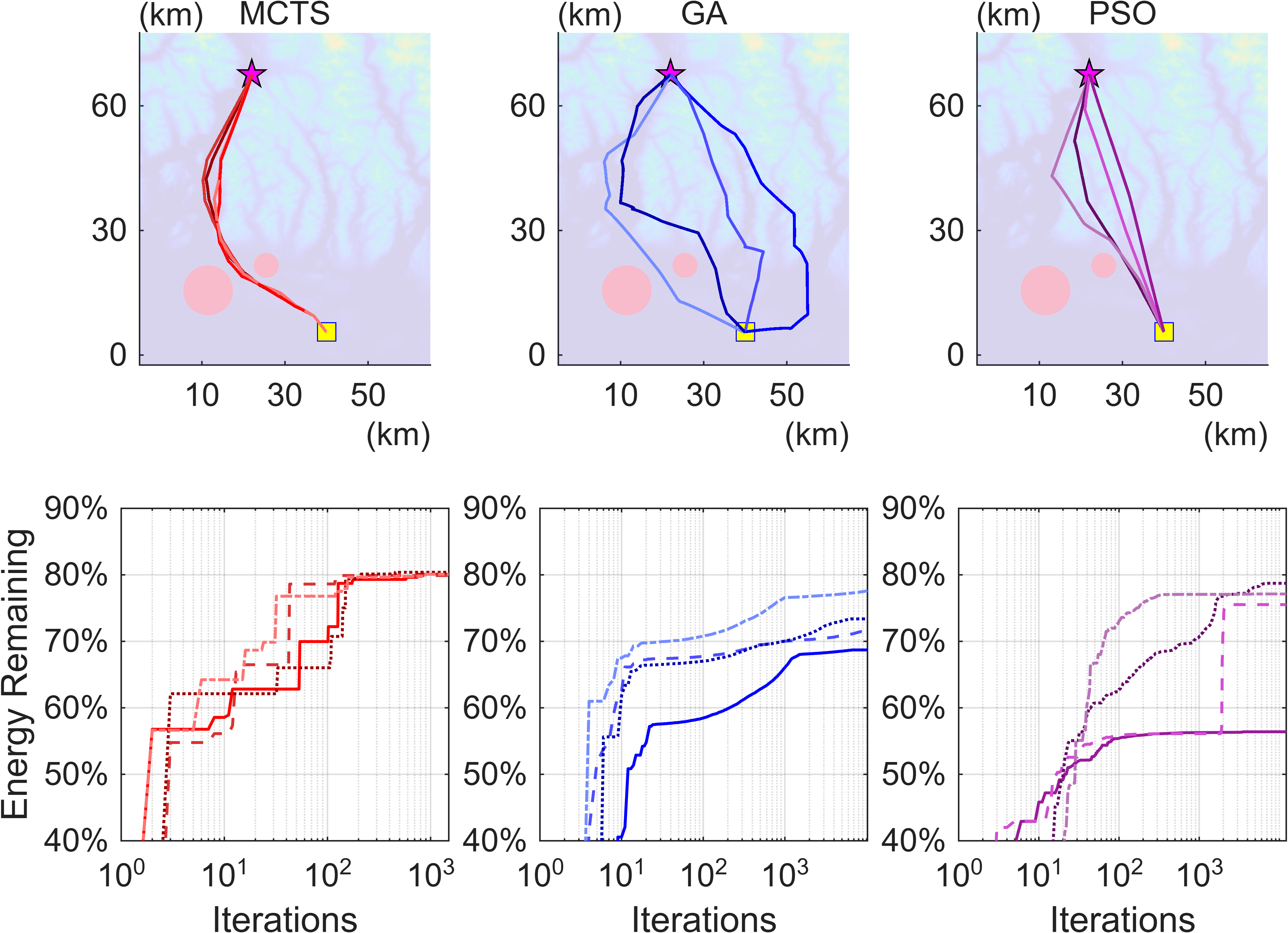}
    \caption{ 
    \change{Seed sensitivity across four random seeds. Top row: planned trajectories for MCTS, GA, and PSO. Bottom row: corresponding reward convergence curves. MCTS consistently discovers the coastal corridor regardless of initialization, whereas GA and PSO exhibit significant route variability across seeds.}
    }
\label{fig:case1_seedSensitivity}
\end{figure}

\change{To assess reliability, we repeated the Case~1 comparison across four runs with randomly drawn initialization biases (Fig.~\ref{fig:case1_seedSensitivity}). MCTS terminal energy remains tightly clustered across all runs ($79.9$--$80.4\%$, std~$0.2\%$), indicating that the tree search converges reliably regardless of initialization. The GA and PSO are considerably more sensitive: GA terminal energy spans $68.7$--$77.6\%$ (std~$3.7\%$) and PSO spans $56.4$--$78.7\%$ (std~$10.4\%$), with both methods strongly influenced by the direction of the initialization bias.}

\paragraph{Summary} These results demonstrate that incorporating environmental forecasts within the planning process enables the generation of energy-efficient and feasible flight plans over long horizons. All four methods were able to find feasible paths but we found that: a) BIT* is efficient and effective as a geometric baseline but cannot exploit wind; \change{b) the GA and PSO readily found feasible solutions but were sensitive to initial conditions and sample-inefficient, and were ultimately outperformed by MCTS in both solution quality and computational efficiency; and c) MCTS reliably converged to a near-optimal solution regardless of its initial guess.} Overall, MCTS outperformed all three baseline methods.

\subsection{Case 2: Online Re-planning under Emerging No-Fly Constraints}

In coastal British Columbia, fast-moving Pacific storm fronts periodically sweep inland. Their forecasts are often uncertain and evolve rapidly; when a front reaches shore, it forms a curved, wave-like boundary that propagates across terrain, leaving behind a broad region of severe turbulence and strong downdrafts. This region is treated as a temporary no-fly zone, and all aircraft must avoid entering it~\citep{navcanada_bcweather}. Therefore, this scenario evaluates the adaptability of the MCTS-based planner when unexpected large-scale weather hazards arise during flight. The objective is to assess online re-planning performance under dynamic constraints.

For this case (Case 2), the initial flight plan is generated without any storm activity. The UAV receives an updated forecast at $t = 300\,\mathrm{s}$ after takeoff indicating that a storm front is approaching. The front is modeled as a sweeping curved boundary that advances inland on a bearing of $60^\circ$ at $90\,\mathrm{km/h}$, generating an expanding no-fly region as it moves. The mission setup is otherwise identical to Case~1 (Sec.~\ref{ssec:case1}, wind profile~1). Because the storm-front information is unavailable during pre-flight planning, both the nominal trajectory and early in-flight refinements are computed without this hazard. Once the updated storm alert is received, the onboard MCTS planner performs online re-planning to generate a feasible diversion path. The focus is on illustrating the benefits of real-time re-planning with MCTS, so the BIT*/GA/PSO baselines are not re-run in this case. \change{The onboard re-planner is allotted a time-based budget in each cycle: MCTS searches from the current state for the wall-clock duration corresponding to the aircraft's flight time to the next waypoint. The number of iterations per cycle is therefore an observed quantity, set by the real time available between successive waypoint updates rather than chosen a-priori. This reflects the operational constraint that re-planning must complete before the vehicle reaches the waypoint at which the new plan takes effect.}

\begin{figure}[h!]
    \centering
    \includegraphics[width=0.48\textwidth]{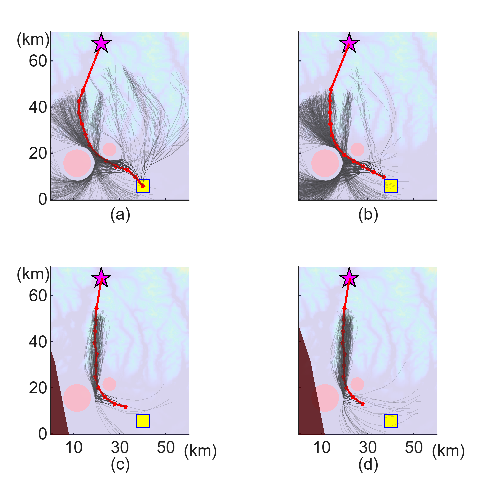}
    \caption{Online MCTS re-planning under an emerging storm front. Gray curves show sampled trajectories, red shows the current flight plan, and crimson-red indicates the storm front. The evolving disturbance induces mid-mission replanning. Snapshots at (a)~$t=0~\mathrm{s}$; (b)~$197~\mathrm{s}$; (c)~$382~\mathrm{s}$; (d)~$598~\mathrm{s}$.}
    \label{fig:case2_allpath}
\end{figure}

Figure~\ref{fig:case2_allpath} illustrates the resulting re-planning sequence under an emerging storm front. Fig.~\ref{fig:case2_allpath}a shows the original flight plan. As the storm front approaches (from the west-south-west), the first in-flight refinement occurs resulting in the trajectory in Fig.~\ref{fig:case2_allpath}b, yielding only a minor update. As the storm front continues to approach, the nominal trajectory becomes infeasible. Within the next re-planning cycle, the MCTS planner reinitializes the search tree and generates a safe diversion (Fig.~\ref{fig:case2_allpath}c). Subsequent cycle (Fig.~\ref{fig:case2_allpath}d) progressively improves the solution, reflecting the anytime nature of the re-planner: as more computation becomes available in each cycle, the trajectory converges toward a \change{near-optimal} route that skirts the moving hazard while preserving safety margins.

\begin{figure}[h!]
    \centering
    \includegraphics[width=0.45\textwidth]{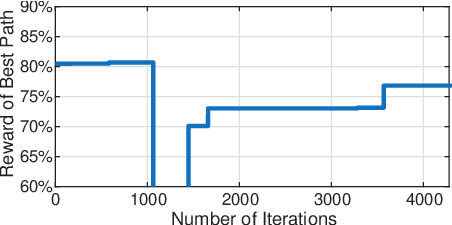}
    \caption{Cumulative reward as a function of MCTS iterations for Case~2.}
    \label{fig:case2_convergence}
\end{figure}

\begin{figure}[h!]
    \centering
    \includegraphics[width=0.45\textwidth]{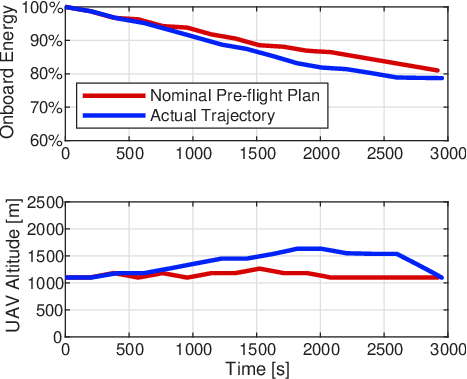}
    \caption{
    Energy and altitude profiles for Case~2: Onboard energy consumption for the actual trajectory vs\ the initial flight plan; Altitude evolution for the actual trajectory vs\ the initial flight plan.
    }
    \label{fig:case2_performance}
\end{figure}

Figures~\ref{fig:case2_convergence} and~\ref{fig:case2_performance} illustrate the computational efficiency and quantitative performance of the re-planner. When the storm front emerges and invalidates the nominal route, a new search tree is initialized from the current aircraft state. In subsequent cycles, valid subtrees from prior searches are retained and refined, accelerating convergence toward \change{near-optimal} diversion paths under the fixed computational budget. Figure~\ref{fig:case2_convergence} shows the evolution of cumulative reward, which reflects the expected remaining energy at mission completion. Before the storm alert, the reward increases steadily. Once the hazard becomes active, the reward drops as the original plan becomes infeasible and the immediate re-plan prioritizes collision avoidance over optimality. Within a few cycles, the planner rapidly recovers most of the lost performance with subtree reuse, achieving a cumulative reward comparable to the pre-storm baseline. Figure~\ref{fig:case2_performance} shows that the actual trajectory incurs a modest increase in energy consumption and avoids aggressive altitude changes throughout the mission.

Throughout the scenario, all re-planned trajectories satisfy terrain and airspace constraints and maintain a safe distance from the storm-affected region. This confirms that the framework preserves operational safety while adapting to rapidly evolving environmental conditions. Overall, this case study demonstrates the robustness and responsiveness of the proposed MCTS-based re-planning framework: with efficient subtree reuse and anytime refinement, the UAV adapts effectively to emergent large-scale weather constraints while maintaining energy efficiency and mission feasibility.

\section{Conclusion}
\label{sec:conclusion}

This paper presented an MCTS-based framework for long-range, energy-aware trajectory planning and online re-planning of fixed-wing UAVs operating in dynamic, constrained environments. By formulating mission planning as a finite-horizon Markov decision process and embedding pre-stabilized vehicle dynamics, wind forecasts, and navigation-field–based constraint handling directly into the search, the proposed method generates dynamically feasible waypoint policies that explicitly optimize remaining onboard energy while accommodating terrain and time-varying airspace restrictions. The simulation results demonstrate that our method systematically exploits wind and terrain structure, achieving substantially higher terminal energy than a BIT* baseline despite modest increases in path length and \change{ proved more efficient, performant, and robust than PSO and GA baselines.} In addition, the closed-loop re-planning architecture enables rapid adaptation to large-scale emergent hazards, producing safe diversion trajectories and recovering near-nominal performance under fixed computational budgets through anytime refinement and subtree reuse. Future work includes incorporating uncertainty about environmental forecasts, reducing the computational footprint of the algorithm, and progression towards experimental validation.

\section*{Acknowledgments}
The authors used ChatGPT (GPT-5, OpenAI) to check grammar during the final preparation of this manuscript. All content was reviewed and verified by the authors, who take full responsibility for the final text.

\section*{Funding Sources}
This work was supported by the Natural Sciences and Engineering Research Council of Canada Discovery Grants Program (Reference \#\,RGPIN-2023-03257).

\bibliography{references}

\end{document}